\documentclass[12pt,draftclsnofoot,onecolumn]{IEEEtran}
\usepackage{array,graphicx,cite,epsfig,amssymb,amsfonts}
\usepackage{bbm,amsmath,epstopdf,algorithm,algorithmic,subfigure, multirow}
\usepackage{comment}
\usepackage{color}
\usepackage{cases}
\usepackage{slashbox}

\newcommand{\be}{\begin{equation}}
\newcommand{\ee}{\end{equation}}
\def\bee#1\eee{\begin{align}#1\end{align}}
\newcommand{\bse}{\begin{subequations}}
\newcommand{\ese}{\end{subequations}}
\newcommand{\nnb}{\nonumber}

\newtheorem{theorem}{\textbf{Theorem}}

\newcommand{\specialcell}[2][c]{%
  \begin{tabular}[#1]{@{}c@{}}#2\end{tabular}}

\ifodd 0
\else

\fi

\begin{document}

\title{Reinforcement Learning for Improved Random Access in Delay-Constrained Heterogeneous Wireless Networks}


\author{Lei~Deng,~\IEEEmembership{Member,~IEEE,}
        Danzhou~Wu,
        Zilong~Liu,~\IEEEmembership{Senior Member,~IEEE,}
        Yijin~Zhang,~\IEEEmembership{Senior Member,~IEEE,}
        and~Yunghsiang~S.~Han,~\IEEEmembership{Fellow,~IEEE}
\thanks{
A preliminary version of this work was presented by IEEE GLOBECOM Workshop on Towards Native-AI Wireless Networks, 2021 \cite{wu2021GC}.
\emph{(Corresponding author: Yunghsiang~S.~Han.)}
}
\thanks{L.~Deng  and D.~Wu are with College of Electronics and Information Engineering,
Shenzhen University, Shenzhen 518060, China
(e-mail: ldeng@szu.edu.cn, wudanzhou2019@email.szu.edu.cn).}
\thanks{Z.~Liu is with School of Computer Science and Electronic Engineering, University of Essex, Colchester CO4 3SQ, U.K. (e-mail:
zilong.liu@essex.ac.uk).}
\thanks{Y.~Zhang is with School of Electronic and Optical Engineering, Nanjing University of Science and Technology, Nanjing 210094, China (e-mail:
yijin.zhang@gmail.com).}
\thanks{Y.~S.~Han is with Shenzhen Institute for Advanced Study, University of Electronic Science and Technology of China, Shenzhen 518110, China  (e-mail: yunghsiangh@gmail.com).}
}


\maketitle


\begin{abstract}
In this paper, we for the first time investigate the random access problem for a delay-constrained  heterogeneous wireless network.
We begin with a simple two-device problem
where two devices deliver delay-constrained traffic to an access point (AP) via a common unreliable collision channel.
By assuming that one device (called Device 1) adopts ALOHA, we aim to optimize
the random access scheme of the other device (called Device 2).
The most intriguing part of this problem is that Device 2 does not know the information
of Device 1 but needs to  maximize the system timely throughput.
We first propose a Markov Decision Process (MDP) formulation
to derive a model-based upper bound so as to quantify the performance gap of certain random access schemes.
We then utilize reinforcement learning (RL) to design an R-learning-based random access scheme, called tiny state-space R-learning random access (TSRA),
which is subsequently extended for the tackling of the general multi-device problem.
We carry out extensive simulations to show that the proposed TSRA simultaneously achieves higher timely throughput, lower computation complexity, and lower power consumption than the existing baseline---deep-reinforcement learning multiple access (DLMA). This indicates that our proposed TSRA scheme is a promising means for efficient random access over massive mobile devices with limited computation and battery capabilities.
\end{abstract}


\IEEEpeerreviewmaketitle

%
%
%
%
%
%
%
%
%
%
%
%
%
%
%

\section{Introduction}\label{sec:introduction}
\IEEEPARstart{W}{ireless} communications are rapidly evolving from connecting people to networking everything in various vertical industries.
Toward that end, hard delay constraint is one of the most important communication requirements in many vertical
applications, such as factory automation, robot collaboration, smart grid load control, autonomous vehicles, online gaming, virtual reality, and tactile Internet, etc. \cite{ts22104,chen2020wireless, gp2015the, kkim2012cyber,lei2017timely,deng2021joint,koutsiamanis2018best, leonardi2019rt}.
In such applications, each packet is assigned a hard delay:
it will expire and then be removed from the system if it has not been delivered successfully within a given delay.
For example, in virtual reality, the motion-to-photon latency is generally at most 15ms; exceeding
this deadline will cause motion sickness and dizziness to the device \cite{elbamby2018toward}.
This hard delay constraint poses an array of challenges to network design, since
many existing techniques designed for legacy communication networks may not work well
for delay-constrained ones \cite{lei2017timely}.

To support various applications in different scenarios, heterogeneous wireless networks are ubiquitous nowadays
where the heterogeneity comes from three aspects. First, massive number of devices
serving a range of applications may have different traffic patterns.
Second, there are a variety of channel conditions to be dealt with in wireless networks, in which the communication quality highly depends on radio propagation conditions such as distance, mobility, shadowing, and fading, etc. 
Third and most importantly, different communication protocols may need to be simultaneously supported in wireless networks.
It is common that different networks, such as cellular (NB-IoT), LoRa,
WiFi, Bluetooth, Zigbee, NFC, co-exist in certain area to deliver data traffic.
To avoid/mitigate interference, most of the current spectrum bands are exclusively assigned among different networks.
Such an exclusively-assigning scheme, however, is hard to
satisfy the explosively increasing wireless traffic, since it is unable to dynamically match the supply and demand.
To address this issue, the Defense Advanced Research Projects Agency (DARPA)
envisions that spectrum should be dynamically and collaboratively shared by
heterogeneous wireless networks. To validate this new spectrum sharing scheme,
DARPA hosted a three-year competition, called Spectrum Collaboration Challenge (SC2),
where teams need to design clean-slate radio techniques to share spectrum with their competitors but without knowing
protocol details of competitors, \emph{with the ultimate goal of increasing overall data throughput} \cite{SC2,tilghman2019will}.
The competition has demonstrated that indeed the new collaboratively-sharing scheme can transmit far more data.
To realize DARPA's vision, we need to re-design PHY, MAC and network layers of wireless networks.
In this paper, we only focus on the MAC layer design, in particular, on the uplink random access scheme design, for heterogeneous wireless networks.

Several random access schemes have been designed in heterogeneous wireless networks under the delay-unconstrained settings.
Yu \emph{et al}. in \cite{yiding2019deep} introduced deep reinforcement learning (DRL) into the random access scheme design for heterogeneous wireless networking.
Their proposed scheme, called deep-reinforcement learning multiple access (DLMA), adopted feedforward neural networks (FNN) as the deep neural network.
In \cite{yiding2020non}, the authors further applied DRL into CSMA and designed a new CSMA variant, called CS-DLMA, for heterogeneous wireless networking.
As compared with DLMA, CS-DLMA adopts recurrent neural networks (RNN) for a non-uniform time-step deep Q-network (DQN) by leveraging the fact that
the time duration required for carrier sensing is smaller than the duration of data transmission. Both \cite{yiding2019deep} and \cite{yiding2020non}
assume a saturated delay-unconstrained traffic pattern.
On the other hand, some works studied random access schemes for delay-constrained communication in homogeneous wireless networking.
Deng \emph{et al}. in \cite{lei2018on} analyzed the asymptotic performance of ALOHA system for frame-synchronized delay-constrained traffic pattern.
Zhang \emph{et al}. in \cite{zhang2019achieving} studied the system throughput and optimal retransmission probability of ALOHA for the saturated delay-constrained traffic.
Campolo \emph{et al}. in \cite{campolo2011modeling} analyzed $p$-persistent CSMA for broadcasting delay-constrained traffic.
An online-learning-based multiple access scheme, called Learn2MAC, was proposed in \cite{destounis2019learn2mac} to provide delay guarantee and low energy consumption.
However, to the best of our knowledge, there have been no works designing uplink random access scheme for delay-constrained heterogeneous wireless  networks.

\begin{table}
\caption{A comparison of our proposed TSRA scheme with existing baselines. \label{tab:compare-4-schemes}}
\centering
\scriptsize
\begin{tabular}{|c|c|c|c|c|}
  \hline
  \textbf{Random Access Scheme} & \textbf{DLMA} \cite{yiding2019deep} & \textbf{Learn2MAC} \cite{destounis2019learn2mac} & \textbf{ALOHA} \cite{lei2018on} & \textbf{TSRA} (This work) \\ \hline
  Timely Throughput & Medium & Low & Low & High \\ \hline
  Computation Complexity & High & Medium & Low & Low \\ \hline
  Power Consumption & High & High & Low & Low \\ \hline
  Methodology & Deep Reinforcement Learning & Online Learning & Simple Stochastic Model & R-Learning \\
  \hline
\end{tabular}
\end{table}

In this paper, we take a first step to fill this blank by designing a reinforcement-learning-based random access scheme, called
\textbf{T}iny \textbf{S}tate-space \textbf{R}-learning random \textbf{A}ccess (TSRA),
which delivers delay-constrained traffic in a heterogeneous wireless network.
We show that TSRA simultaneously achieves higher timely throughput, lower computation complexity, and lower power consumption
than existing baseline schemes, including DLMA \cite{yiding2019deep} and Learn2MAC \cite{destounis2019learn2mac},
for a large number of settings with up to 100 devices.
A comparison of our proposed TSRA with three existing baseline schemes is  shown in Table~\ref{tab:compare-4-schemes}.
As we can see, the proposed TSRA represents a promising enabling approach for efficient wireless networking where the devices may suffer from limited computation and battery capabilities.

The key idea of TSRA originates from a comprehensive analysis for a simple two-device problem
where there are only two devices accessing a wireless channel for delivering delay-constrained traffic.
Whilst one device (called Device 1) adopts the slotted ALOHA scheme, we optimize the random access scheme of
the other (called Device 2) with the goal of maximizing the system timely throughput.
Assuming that Device 2 does not know the information of Device 1, the primary objective here is how to maximize the system timely throughput. The analysis and proposed random access scheme
for the simple two-device case can be naturally extended to the general multi-device case. Thus, in the rest of this paper,
we will concentrate on such a two-device problem before describing the general multi-device problem.
Specifically, our major contributions of this paper are summarized as follows:
\begin{itemize}
\item For the two-device case, we first establish a  model-based  upper bound, and then propose an average-reward model-free R-learning-based random access scheme \cite{schwartz1993reinforcement,singh1994reinforcement,sutton2018reinforcement}.
We illustrate that the average-reward-based R-learning is more effective than the widely-used discounted-award-based Q-learning for our problem.
Since the state space of R-learning exponentially increases with $D$,
we further exploit the problem structure and design the low-complexity TSRA scheme by only utilizing the information about
whether Device 2 has a most urgent packet (which will expire in one slot).
\item We further extend TSRA to the general multi-device case by carefully re-designing the reward function.
Since there are significantly more system possibilities in the multi-device case,
we show that the reward function with only two values in the two-device case is not suitable for the general multi-device case.
Instead, we design a more sophisticated reward function with four value levels to better differentiate different system possibilities.
\item Finally, we conduct extensive simulations to demonstrate the superior performance of TSRA.
For the two-device case, we show that the system timely throughput of TSRA is $5.62\%$ higher than that of the existing baseline
DLMA \cite{yiding2019deep} and is only 4.98\% lower than the derived upper bound.
The time and space complexity of TSRA are respectively 96.36\% and 83.62\% lower than  that of DLMA.
For the multi-device case, TSRA increases the system timely throughput by 79.19\% than DLMA,
and the time and space complexity of TSRA are respectively 98.29\% and 90.74\% lower than that of DLMA.
Furthermore, TSRA reduces the system power consumption by 52.29\% as compared to DLMA.
\end{itemize}

The rest of this paper is organized as follows.  In Secs.~\ref{sec:system_model}-\ref{sec:TSRA},
we study the two-device problem in great detail and illustrate how we design the TSRA scheme.
In particular, for the two-device problem, we describe the system model in Sec.~\ref{sec:system_model},
derive an MDP-based upper bound for the performance  in Sec.~\ref{sec:upper_bound},
and propose the TSRA scheme in Sec.~\ref{sec:TSRA}.  In Sec. V, we extend TSRA to the general multi-device problem. In Sec. VI, we carry
out extensive simulations to compare TSRA with existing baselines. Sec. VII concludes this paper.

\section{System Model and Problem Formulation: A Two-Device Case}\label{sec:system_model}
\begin{figure}[t]
  \centering
  \includegraphics[width=0.7\linewidth]{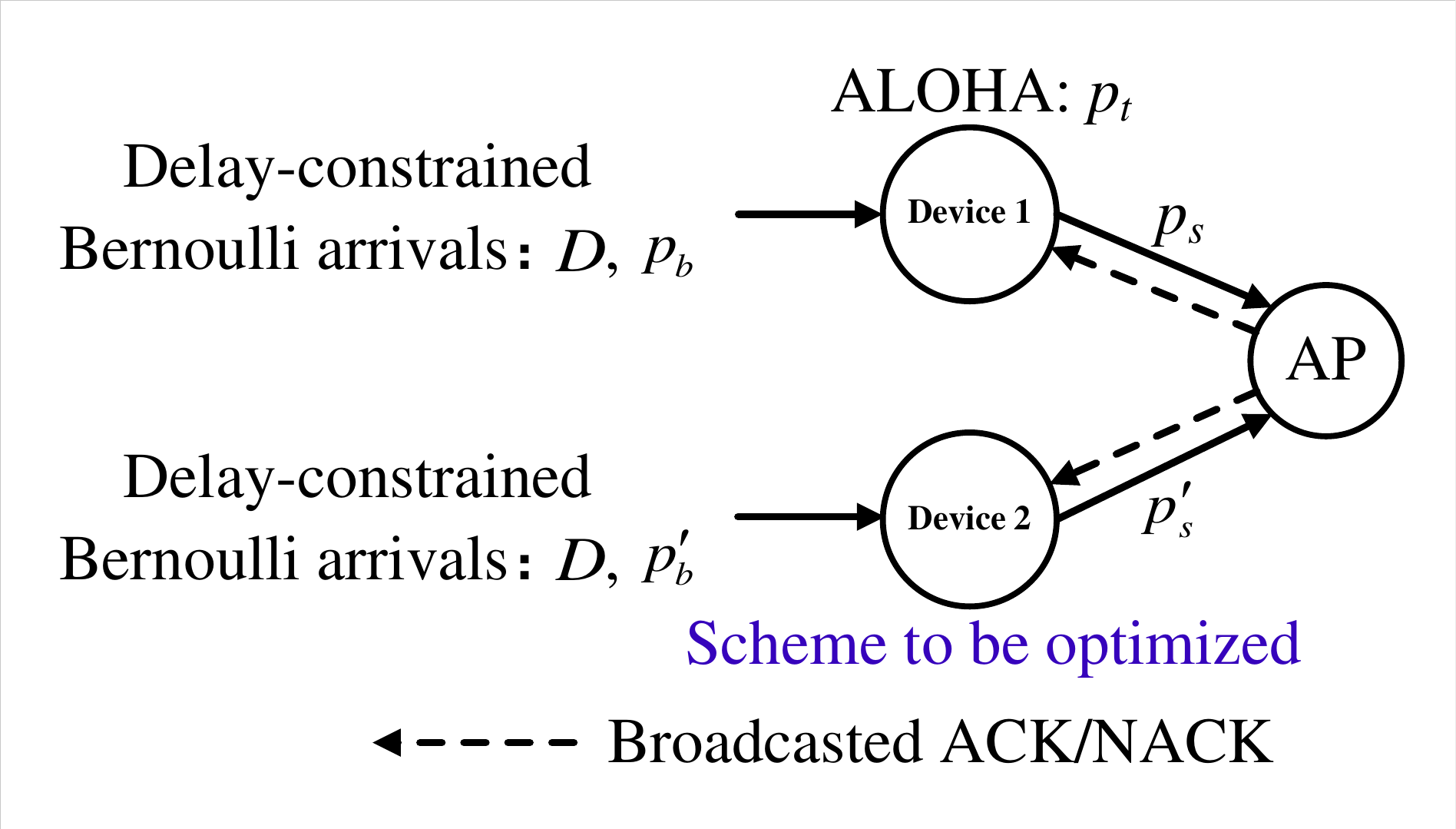}
  \caption{System model for the two-device case.}\label{fig:system_model}
\end{figure}

As we mentioned in the previous section, our proposed TSRA scheme is fundamentally derived from a simple two-device case.
We thus first describe the system model and problem formulation for the two-device case in this section.
The system model is depicted in Fig.~\ref{fig:system_model}.
Specifically, two devices share a wireless channel to
deliver delay-constrained traffic to an access point (AP). Time is slotted and indexed
from slot 1. A slot spans the time duration for an device to transmit a packet to the AP and get a feedback from the AP.
For easier reference, we use ``at slot $t$" to refer to ``at the beginning of slot $t$" and
use ``in slot $t$" to refer to ``in the time space of slot $t$".
System state changes at a slot and transmission behaviour occurs in a slot.
To facilitate the analysis, we first assume a delay-constrained Bernoulli traffic pattern for both devices: Device 1 (resp. Device 2)
has a new packet arrival with  probability $p_b \in (0,1]$ (resp. $p'_b \in (0,1]$) at any slot, and
all packets have a hard delay of $D$ slots. Note that the Bernoulli arrivals characterized by an arrival rate $p_b$ or $p_b'$
are widely used in wireless communication research, e.g., \cite{dua2008random, sun2019closed, yang2021understanding}.
A packet will be removed from the system
if it has not been delivered successfully to the AP in $D$ slots after its arrival. We will later investigate different traffic patterns
in Sec.~\ref{sec:simulation}.

We assume an \emph{unreliable collision} wireless channel model. If both devices transmit a packet to the AP in a slot,
then a channel collision occurs and both packets cannot be successfully received/decoded by the AP. Even though only one device transmits a packet
to the AP, the wireless channel is still unreliable due to shadowing and fading. We model such unreliability by a
success probability. Specifically, if only Device 1 (resp. Device 2) transmits a packet to the AP, the packet
can be successfully delivered with probability $p_s \in (0,1]$ (resp. $p'_s \in (0,1]$). Otherwise,  a channel
error happens. Thus, a transmission failure may occur either due to a channel collision or due to a channel error.
Note the we model an unreliable channel by a parameter  $p_s$ or $p_s'$, which is also widely used
in delay-constrained wireless communications, e.g., \cite{hou2009qos,hou2010scheduling,hou2010utility,lei2017timely}.

The two-device network is heterogeneous, not only because they could have different arrival probabilities and different channel success probabilities,
but also because they may use different random access schemes (i.e., transmission policies).
We assume that Device 1 adopts the slotted ALOHA\footnote{For simplicity, we will use ``ALOHA" to represent ``slotted ALOHA" in the rest of this paper.} protocol with transmission/retransmission probability $p_t \in [0,1]$.
Namely, Device 1 always transmits or retransmits its head-of-line (HoL) packet to the AP with probability $p_t$ in any slot.
The random access scheme of Device 2 is under our control. We design its random access scheme $\pi$
so as to maximize the system timely throughput \cite{lei2018on}, which is defined as
\bee
R^{\pi} \triangleq \liminf_{T \to \infty} \frac{\mathbb{E}^\pi \left[ \substack{\text{\# of packets of both devices delivered successfully}  \text{ before expiration from slot $1$ to slot $T$}} \right]}{T}.
\label{equ:def-R}
\eee
The expectation is taken over all system randomness and possibly policy randomness.
Note that our design space is Device 2's scheme while our goal is to maximize the system timely throughput.
This is in line with DARPA's vision on collaboratively-sharing scheme for spectrum  \cite{SC2,tilghman2019will}.

Note that both devices cannot communicate with each other. Thus, Device 2 does not know the queue information and transmission information of Device 1.
However, it can interact with the environment (i.e., the AP) to learn the information of Device 1.
Specifically, at the end of a slot, the AP will broadcast an acknowledgement (ACK) to both devices if it successfully decodes a packet,
broadcast a negative-acknowledgement  (NACK) if it receives at least one packet but does not decode it successfully (either due to a channel collision or due to a channel error),
and broadcast nothing if it does not receive any packet in this slot.
By observing such feedback, Device 2 aims to infer the behaviour of Device 1
and then judiciously design its access scheme. This motivates us to use reinforcement learning (RL) \cite{sutton2018reinforcement}
due to its great success in solving many interactive decision problems in a model-free manner.

\section{A Model-based Upper Bound for the Two-Device Case} \label{sec:upper_bound}
Before we present our model-free RL-based random access scheme in Sec.~\ref{sec:TSRA},
we present a model-based upper bound for the two-device case in this section. Suppose
that Device 2 is aware of Device 1's parameters including $p_b$, $D$, $p_t$, and $p_s$,
and queue information (i.e., the number of packets in the queue and the arrival time of each packet) at the beginning of any slot.
However, when Device 2 decides to transmit or not transmit a packet in any slot,
it cannot know whether Device 1 transmits/retransmits a packet or not in the same slot.
Otherwise, Device 2 can always avoid collision such that the problem becomes trivial.
Such a model-based setting with more revealed information to Device 2
allows us to derive an upper bound for the system timely throughput of the original model-free problem.
This upper bound will serve as a performance benchmark for evaluating any random access policy later,
as we can numerically gauge the performance gap.

We first consider the special case of the hard delay $D=1$. Note that
$D=1$ means that any packet arriving at (the beginning of) slot $t$
will expire at the end of slot $t$ (i.e., at the beginning of slot $t+1$) if it is not transmitted or its transmission fails
due to channel collision or channel error. Thus, at any slot,
the queue of both devices has at most one packet. This significantly simplifies
the system design due to the coupling-free feature between different slots. We can thus
derive the optimal policy of Device 2, which surprisingly is a binary policy, as shown in the following theorem.

\begin{theorem} \label{thm:D=1}
If $D=1$, then the optimal strategy of Device 2 to maximize the system timely throughput is i) always transmitting the HoL packet if
\be
p_bp_t < \frac{p'_s}{p_s + p'_s}, \label{equ:always-transmit-D=1}
\ee
and ii) always remaining idle if its queue is empty or \eqref{equ:always-transmit-D=1} does not hold.
The corresponding system timely throughput is
\be
R=
\left\{
  \begin{array}{ll}
    \left[ p'_{s}-(p_{s}+p'_{s})p_{t}p_{b}\right]p'_{b}+p_{s}p_{t}p_{b}, & \hbox{if \eqref{equ:always-transmit-D=1} holds;} \\
    p_{s}p_{t}p_{b}, & \hbox{otherwise.}
  \end{array}
\right.
\label{equ:upper-bound-D=1}
\ee
\end{theorem}
\begin{IEEEproof}
Please see the proof in Appendix~\ref{app:proof-of-theorem-D=1}.
\end{IEEEproof}

Let us investigate condition \eqref{equ:always-transmit-D=1} by assuming that $p_s=p'_s$, i.e., both devices have the same channel quality.
In this case, condition \eqref{equ:always-transmit-D=1} becomes $p_bp_t < 0.5$. Note that $p_b p_t$ is
the probability that Device 1 transmits a packet in any slot since $D=1$. Thus, if this probability is less than 0.5, i.e., Device 1 is not aggressive,
Device 2 will become completely aggressive to take over the channel.
Conversely, if this probability is larger than 0.5, i.e., Device 1 is aggressive,
Device 2 will become completely unaggressive to hand over the channel.
Such a binary policy achieves the best tradeoff between utilizing the wireless channel
and avoiding collision. The closed-form expression \eqref{equ:upper-bound-D=1} serves as an upper bound for $D=1$.

However, for general $D>1$, it is difficult to directly characterize an optimal strategy and the corresponding system timely throughput
because of the coupling feature between different slots. Since both the Bernoulli traffic pattern and the collision channel are memoryless 
and the queue dynamics only depend on its current state, our system preserve the Markov property. 
We thus formulate our problem as an MDP problem
and propose an upper bound by numerically solving a linear programming problem. An MDP is characterized by its state, action, reward function, and transition probability \cite{puterman1994markov}.
The state of the system at (the beginning of) slot $t$ is defined as
\be
s_t \triangleq (l_{t,1}, l_{t,2}, o_t). \label{equ:S1}
\ee
In \eqref{equ:S1}, $l_{t,i}=(l^1_{t,i},l^2_{t,i},\cdots,l^D_{t,i})$ is the lead time vector \cite{lei2017timely} of Device $i \in \{1,2\}$ at slot $t$, where
\be
l^k_{t,i} =
\left\{
  \begin{array}{ll}
    1, & \hbox{if Device $i$ has a packet at slot $t$, which will expire in $k \in \{1,2,\cdots,D\}$ slots;} \\
    0, & \hbox{otherwise.}
  \end{array}
\right.
\nnb
\ee
Further, $o_t \in \{\textsf{BUSY}, \textsf{SUCCESSFUL}, \textsf{IDLE}, \textsf{FAILED}\}$ is the channel observation at (the beginning of) slot $t$,
equivalently, at the end of slot $t-1$.
Depending on whether devices transmit a packet in slot $t-1$,
the interpretation of these four different observations is shown in Table ~\ref{tab:observation-two-devices}.
Specifically,
channel observation $o_t=\textsf{IDLE}$ means that Device 2 receives nothing from the AP at the end of slot $t-1$,
indicating that both devices do not transmit a packet in slot $t-1$.
Channel observation $o_t=\textsf{BUSY}$
means that Device 2 does not transmit a packet but receives an ACK from the AP in slot $t-1$,
indicating that Device 1 transmits a packet and no channel error happens for Device 1's packet.
Channel observation $o_t=\textsf{SUCCESSFUL}$ means that Device 2 transmits a packet
and receives an ACK from the AP in slot $t-1$, indicating that
Device 1 does not transmit a packet and no channel error happens for Device 2's packet.
Channel observation $o_t=\textsf{FAILED}$ means that Device 2 receives an NACK from the AP in slot $t-1$.
This is the most complicated observation and it has two different interpretations. If Device 2 does not transmit a packet in slot $t-1$,
this observation means that Device 1 transmits a packet and a channel error happens in slot $t-1$.
If Device 2 transmits a packet in slot $t-1$, this observation means that
a channel collision or a channel error happens in slot $t-1$ (Device 2 cannot distinguish these two possibilities).
Without loss of generality, we assume that $o_1=\textsf{IDLE}$.
We remark that the modeling for channel observation  is the same as \cite{yiding2019deep}.
The set of all possible states is denoted by $\mathcal{S}$.
Clearly, we have $\left|\mathcal{S} \right|=2^D \cdot 2^D \cdot 4 = 2^{2D+2}$.

\begin{table}[]
\caption{The interpretation of four different channel observations, depending on whether Device 2 transmits a packet or not. \label{tab:observation-two-devices}}
\scriptsize
\begin{tabular}{|c|c|c|}
\hline
\textbf{Observation} $o_t$ & \textbf{Device 2 transmits a packet in slot $t-1$}    &  \textbf{Device 2 does not transmit a packet in slot $t-1$}                                                                              \\ \hline
\textsf{IDLE}        & Impossible   & Device 1 does not transmit a packet in slot $t-1$                           \\ \hline
\textsf{BUSY}        & Impossible            & \specialcell{Device 1 transmits a packet in slot $t-1$, \\ and no channel error happens for Device 1} \\ \hline
\textsf{SUCCESSFUL}  &  \specialcell{Device 1 does not transmit a packet in slot $t-1$, \\ and no channel error happens for Device 2} & Impossible                                                                                                  \\ \hline
\textsf{FAILED}     &  \specialcell{Device 1 transmits a packet in slot $t-1$ and a channel \\ collision happens,  or
  Device 1 does not transmit a packet \\ and a channel error happens for Device 2}          &
 \specialcell{Device 1 transmits a packet in slot $t-1$, \\ and a channel error happens for Device 1}    \\ \hline
\end{tabular}
\end{table}

In slot $t$, the action of Device 2 is denoted by $a_t$. Similar to \cite{yiding2019deep}, the action space is defined as $\mathcal{A} \triangleq \{\textsf{TRANSMIT}, \textsf{WAIT}\}$.
One can readily prove that it is optimal to first transmit the HoL packet if there are multiple packets in the Device 2's queue at any slot.
Thus, action $a_t=\textsf{TRANSMIT}$ means that Device 2 transmits its HoL packet in slot $t$, while $a_t=\textsf{WAIT}$ means that
Device 2 does not transmit a packet in slot $t$.

We define the reward function $r(s_t,a_t)$ as
\bee
r(s_t,a_t) \triangleq 1_{\left\{o_t \in \{ \textsf{BUSY}, \textsf{SUCCESSFUL}\} \right \}}, \forall s_t \in \mathcal{S}, a_t \in \mathcal{A},
\label{equ:reward function}
\eee
where $1_{\{\cdot\}}$ is the indicator function. Note that $o_t=\textsf{BUSY}$
means that Device 1 transmits a packet successfully in slot $t-1$,
and $o_t= \textsf{SUCCESSFUL}$ means that Device 2 transmits a packet successfully in slot $t-1$.
Thus $r(s_t,a_t)=1$ if the system (either Device 1 or Device 2) transmits a packet successfully in slot $t-1$.
Note that we model the reward with ``delay of gratification". Namely,
the delivered packet in slot $t-1$ is translated into the reward at slot $t$.
However, since our performance metric is long-term system timely throughput,
such ``delay of gratification" will not cause performance loss, as shown in \eqref{equ:R-is-same-as-avg-reward} later.
In addition, we remark that the reward function only depends on the channel observation $o_t$, regardless of queue information $l_{t,1}$ and $l_{t,2}$.

The transition probability from state $s$ at slot $t$ to state $s'$ at slot $t+1$ if taking action $a$ in slot $t$ is defined as
\be
P(s'|s,a) \triangleq P(s_{t+1}=s'|s_{t}=s, a_t=a), \forall t, s,s',a, \label{equ:MDP-P}
\ee
which depends on
(i) the arrival and expiration events of both devices,
(ii) the transmission events of both devices,
(iii) the channel collision and channel error events,
and (iv) the change of lead time vector.
We use an example to illustrate how to compute the transition probabilities;
see \textsl{https://github.com/DanzhouWu/TSRA/tree/main/two\_device/TransitionProbaility}.

Based on the above MDP model, it is straightforward to see that the system timely throughput under a policy $\pi$ defined in \eqref{equ:def-R} is equivalent
to the average reward of our formulated MDP under policy $\pi$, i.e.,
\bee
R^{\pi} & = \liminf_{T \to \infty} \frac{\mathbb{E}^\pi \left[ \substack{\text{\# of packets of both devices delivered successfully}  \text{ before expiration from slot $1$ to slot $T$}} \right]}{T}, \nnb \\
 & =  \liminf_{{T}\rightarrow\infty} \frac{\sum_{t=2}^{T+1} \mathbb{E}^{\pi}\{r(s_t, a_t)\} }{{T}}, \nnb \\
 & =  \liminf_{{T}\rightarrow\infty} \frac{\sum_{t=1}^{T} \mathbb{E}^{\pi}\{r(s_t, a_t)\} }{{T}}. \label{equ:R-is-same-as-avg-reward}
\eee
Again, note that the expectation is taken over all system randomness and possibly policy randomness.
Thus, our problem becomes an average-reward MDP problem.
Here we use the dual linear program approach to solve this MDP problem \cite[Chapter 9.3]{puterman1994markov},
\bee
\max & \quad \sum_{s \in \mathcal{S}} \sum_{a \in \mathcal{A}} r(s, a) x(s,a) \nnb \\
\text{s.t.} & \quad \sum_{a \in \mathcal{A}} x(s',a) = \sum_{s \in \mathcal{S}} \sum_{a \in \mathcal{A}} P(s'|s, a) x(s,a), \quad \forall s' \in \mathcal{S} \nnb \\
& \quad \sum_{a \in \mathcal{A}} x(s',a) + \sum_{a \in \mathcal{A}} y(s',a) = \sum_{s \in \mathcal{S}} \sum_{a \in \mathcal{A}} P(s'|s, a) y(s, a) + \alpha_{s'}, \quad \forall s' \in \mathcal{S} \quad \nnb \\
\text{var.} & \quad x(s, a) \geq 0, \quad y(s, a) \geq 0,  \quad \forall s \in \mathcal{S}, a \in \mathcal{A} \label{equ:MDP-LP}
\eee
where $\{\alpha_{s}: s \in \mathcal{S}\}$ are arbitrary constants such that $\alpha_s > 0 \;\; (\forall s \in \mathcal{S})$ and $\sum_{s \in\mathcal{S}}\alpha_{s} = 1$. Note that we follow standard procedures of the dual linear program
 in \cite[Chapter 9.3]{puterman1994markov}. Basically, notation
$x(s,a)$ (resp. $y(s,a)$) represents the frequency (or the stationary probability) that
the Markov chain is on state $s$ and the action is $a$ where $s$ is a recurrent state (resp. a transient state);
see \cite[Proposition 9.3.2]{puterman1994markov}.

The optimal value of problem \eqref{equ:MDP-LP} serves as an upper bound for the system timely throughput for general $D>1$. In addition, according to \cite{hordijk1979linear} and \cite[Chapter 9.3.1]{puterman1994markov},
solving problem \eqref{equ:MDP-LP} also yields a randomized optimal policy,
\bee
\pi(a|s) =
\left\{
  \begin{array}{ll}
    \frac{x^*(s,a)}{\sum_{a\in \mathcal{A}} x^*(s,a)}, & \hbox{if $\sum_{a\in \mathcal{A}}{x^*(s,a)} > 0$;} \\
    \frac{y^*(s,a)}{\sum_{a\in \mathcal{A}} y^*(s,a)}, & \hbox{otherwise.}
  \end{array}
\right.
\label{equ:upper-bound-policy}
\eee
where $\pi(a|s)$ is the probability of taking action $a \in \mathcal{A}$ under state $s \in \mathcal{S}$,
and $\{(x^*(s,a), y^*(s,a)): s \in \mathcal{S}, a \in \mathcal{A}\}$ is an optimal solution of problem \eqref{equ:MDP-LP}.

\section{Tiny State-space R-learning Random Access (TSRA) Scheme for the Two-Device Case}\label{sec:TSRA}

The disadvantage of model-based MDP is that Device 2 needs to know Device 1's parameters and queue information. However, such information cannot
be obtained in practice such that Device 2 cannot know Device 1's queue state $l_{t,1}$ and
the transition probabilities $P(s'|s,a)$ (Please refer to \eqref{equ:S1} and \eqref{equ:MDP-P}).
To address this issue, reinforcement learning (RL) has been proposed as a model-free approach to solve MDP problems.
RL needs the state space $\mathcal{S}$, the action space $\mathcal{A}$, and the reward function $r(s,a), \forall s \in \mathcal{S}, a \in \mathcal{A}$,
but does not need the transition probabilities $P(s'|s,a)$ of an MDP. Instead, RL learns the model by directly interacting with the environment.

Since Device 2 cannot know Device 1's queue information, we define its state at slot $t$ as\footnote{With a little bit abuse of notation,
in the model-free problem in this section, except for the state space, we adopt the same notations of the model-based problem in Sec.~\ref{sec:upper_bound}.
Namely, we still use $s_t$ to denote
the state, $a_t$ to denote the action, and $r(s_t,a_t)$ to denote the reward function for the model-free problem in this section.
They are distinguishable in the context.},
\be
s_t \triangleq (l_{t,2}, o_t), \label{equ:equ-state-FSQA-and-FSRA}
\ee
where $l_{t,2}$ is the queue information of Device 2 itself, and $o_t$ is the channel observation (same as Sec.~\ref{sec:upper_bound}).
The state space $\mathcal{S}'$ is thus of size $2^D \cdot 4 = 2^{D+2}$.
The action space  $\mathcal{A} = \{\textsf{TRANSMIT}, \textsf{WAIT}\}$ is again the same as that in Sec.~\ref{sec:upper_bound}.
The reward function $r(s_t,a_t)$ is defined as
\bee
r(s_t,a_t) \triangleq 1_{\left\{o_t \in \{ \textsf{BUSY}, \textsf{SUCCESSFUL}\} \right \}}, \forall s_t \in \mathcal{S}', a_t \in \mathcal{A},
\label{equ:reward function_RL}
\eee
which is similar to that in the model-based setting (Please refer to \eqref{equ:reward function}).
Namely, the reward is 1 if Device 2 receives an ACK, either for its own packet ($o_t=\textsf{SUCCESSFUL}$) or for Device 1's packet ($o_t=\textsf{BUSY}$).
Otherwise, the reward is 0.

\subsection{Q-Learning} \label{subsec:Q-learning}
Based on the above information, we can apply different RL methods to solve our problem in a model-free manner,
such as Monte Carlo and temporal-difference learning \cite{sutton2018reinforcement}.
Among them, Q-learning is one of the most widely-used methods \cite{sutton2018reinforcement}.
In fact, the delay-unconstrained counterpart of our problem, i.e., \cite{yiding2019deep}, also used Q-learning.
The simplest form of Q-learning, called one-step Q-learning, iteratively updates the Q-function $Q(s,a)$ as follows,
\bee
Q(s_t, a_t) & \leftarrow Q (s_t, a_t) + \alpha \big[r(s_t, a_t) +  \gamma \max_{a \in \mathcal{A} }Q(s_{t+1}, a) - Q(s_t, a_t) \big], \label{equ:Q-learning-Q-function}
\eee
where $\alpha \in (0,1]$ is the learning rate, $\gamma \in (0,1)$ is the discount factor, and
$Q(s,a)$ is the state-action value function (called Q-function),  approximating the discounted reward for given state and action
for the iteratively updated policy $\pi$, i.e.,
\be
Q(s,a) \approx \mathbb{E}^{\pi} \left[\sum_{\tau=t}^{\infty} \gamma^{\tau-t}r(s_\tau, a_\tau) | s_t=s, a_t=a \right]. \label{equ:Q-function-approx}
\ee
Note that the policy $\pi$ is iteratively updated by selecting action $a$ to maximize $Q(s,a)$ for any state $s$
with an $\epsilon$-greedy algorithm \cite{sutton2018reinforcement}.
We call the algorithm \textbf{F}ull \textbf{S}tate-space \textbf{Q}-learning random \textbf{A}ccess (FSQA),
which is detailed in Algorithm~\ref{alg:ql}.

Q-learning is suitable for solving MDPs with discounted reward in a model-free manner.
However, in network communication research, the major performance metric, throughput or timely throughput, is a long-term average reward.
Therefore, Q-learning may be less suitable for network communication research than another RL method, called R-learning,
which solves MDPs with average reward in a model-free manner \cite{schwartz1993reinforcement, singh1994reinforcement,sutton2018reinforcement}.

\begin{algorithm}[t]
 \caption{FSQA Algorithm for Device 2 in the Two-Device Problem}
 \label{alg:ql}
\begin{algorithmic}[1]
  \STATE Initialize Q-function $Q(s, a) = 0$, $\forall s \in \mathcal{S}'$, $\forall a \in \mathcal{A}$,
    \STATE Set learning rate $\alpha=0.01$
    \STATE Initialize the discount factor $\gamma=0.9$
    \STATE Observe the initial system state $s_1$
    \FOR{$t = 1, 2, \cdots$}
    \STATE Choose $a_t$ with an $\epsilon$-greedy algorithm, i.e.,
    \[
    a_t=
    \left\{
      \begin{array}{ll}
        \arg \max_{a \in \mathcal{A} } Q(s_t, a), & \hbox{with prob. $1-\epsilon_t$;} \\
        \text{random action}, & \hbox{with prob. $\epsilon_t$,}
      \end{array}
    \right.
    \]
    where $\epsilon_t=\max\{0.995^{t-1},0.01\}$
    \STATE Observe $r(s_t, a_t)$, $s_{t+1}$
    \STATE Update Q-function as follows,
    		\bee
	           Q(s_t, a_t) & \leftarrow Q(s_t, a_t) + \alpha \big( r(s_t, a_t)  + \gamma \max_{a \in \mathcal{A} }Q(s_{t+1}, a) - Q(s_t, a_t) \big) \nnb
			\eee
    \ENDFOR
\end{algorithmic}
\end{algorithm}

\subsection{R-Learning} \label{subsec:R-learning}

R-learning also utilizes the state-action value function (we still call it Q-function by convention), which however has a different meaning.
Among the variants of R-learning \cite{schwartz1993reinforcement, singh1994reinforcement,sutton2018reinforcement},
in this paper, we adopt the version in \cite[Algorithm 3]{singh1994reinforcement} and \cite[Figure 11.2]{sutton2018reinforcement},
\bee
Q(s_t, a_t) & \leftarrow Q(s_t, a_t) + \alpha \big[ r(s_t, a_t) + \max_{a \in \mathcal{A}  }Q(s_{t+1}, a) - Q(s_t, a_t) - \rho \big], \label{equ:upgrade_Q}\\
& \rho \leftarrow \rho + \beta\big[ r(s_t, a_t) + \max_{a \in \mathcal{A}  }Q(s_{t+1}, a) - Q(s_t, a_t) - \rho \big], \label{equ:upgrade_rho}
\eee
where $\alpha \in (0,1]$ and $\beta \in (0,1]$ are learning rates, $\rho$ approximates the state-independent
average reward for the iteratively updated policy $\pi$, i.e.,
\be
\rho \approx \lim_{T \to \infty} \mathbb{E}^{\pi} \left[ \frac{\sum_{t=1}^T r(s_t,a_t)}{T} \right],
\label{equ:rho-approx-R-learning}
\ee
and Q-function $Q(s, a)$ approximates the state-dependent cumulative reward difference (called relative value in \cite{singh1994reinforcement,sutton2018reinforcement}) for the iteratively updated policy $\pi$, i.e.,
\be
Q(s,a) \approx  \mathbb{E}^{\pi}  \left[ \left. \sum_{\tau=t}^{\infty}  \left[r(s_\tau,a_\tau) - \rho \right] \right| s_t = s, a_t=a \right].
\label{equ:Q-approx-R-learning}
\ee
Similar to Q-learning, the policy $\pi$ is iteratively updated by selecting action $a$ to maximize $Q(s,a)$ for any state $s$
with an $\epsilon$-greedy algorithm.
We call the algorithm \textbf{F}ull \textbf{S}tate-space \textbf{R}-learning random \textbf{A}ccess (FSRA),
which is detailed in Algorithm~\ref{alg:rl}.

\begin{algorithm}[t]
 \caption{FSRA/HSRA/TSRA Algorithm for Device 2 in the Two-Device Problem}
 \label{alg:rl}
\begin{algorithmic}[1]
    \STATE Initialize Q-function $Q(s, a) = 0, \forall a \in \mathcal{A}$, $\forall s \in \tilde{\mathcal{S}}$, where the state space $\tilde{\mathcal{S}}$
is different for different algorithms,
    \[
    \tilde{\mathcal{S}}=
    \left\{
      \begin{array}{ll}
         \mathcal{S}', & \hbox{If the algorithm is FSRA;} \\
         \mathcal{S}'', & \hbox{If the algorithm is HSRA;} \\
         \mathcal{S}''', & \hbox{If the algorithm is TSRA;} \\
      \end{array}
    \right.
    \]
    \label{line:diff-state-space}
    \STATE Initialize $\rho = 0$
    \STATE Set learning rates $\alpha=0.01$, $\beta=0.01$
    \STATE Observe the initial system state $s_1$
    \FOR{$t = 1, 2, \cdots$}
    \STATE Choose $a_t$ with an $\epsilon$-greedy algorithm, i.e.,
    \[
    a_t=
    \left\{
      \begin{array}{ll}
        \arg \max_{a \in \mathcal{A}  } Q(s_t, a), & \hbox{with prob. $1-\epsilon_t$;} \\
        \text{random action}, & \hbox{with prob. $\epsilon_t$,}
      \end{array}
    \right.
    \]
    where $\epsilon_t=\max\{0.995^{t-1},0.01\}$
    \STATE Observe $r(s_t, a_t)$, $s_{t+1}$
    \STATE Update Q-function as follows,
    		\bee
			Q(s_t, a_t) & \leftarrow Q(s_t, a_t) + \alpha \left[ r(s_t, a_t)  + \max_{a \in \mathcal{A}  }Q(s_{t+1}, a) - Q(s_t, a_t) - \rho \right] \nnb
			\eee
    \STATE Update $\rho$ as follows,
    		\bee
    		\rho & \leftarrow \rho + \beta \left[ r(s_t, a_t) +  \max_{a \in \mathcal{A}  }Q(s_{t+1}, a) - Q(s_t, a_t) - \rho \right] \nnb
    		\eee
    \ENDFOR
\end{algorithmic}
\end{algorithm}

\begin{figure}[t]
  \centering
  \includegraphics[width=0.6\linewidth]{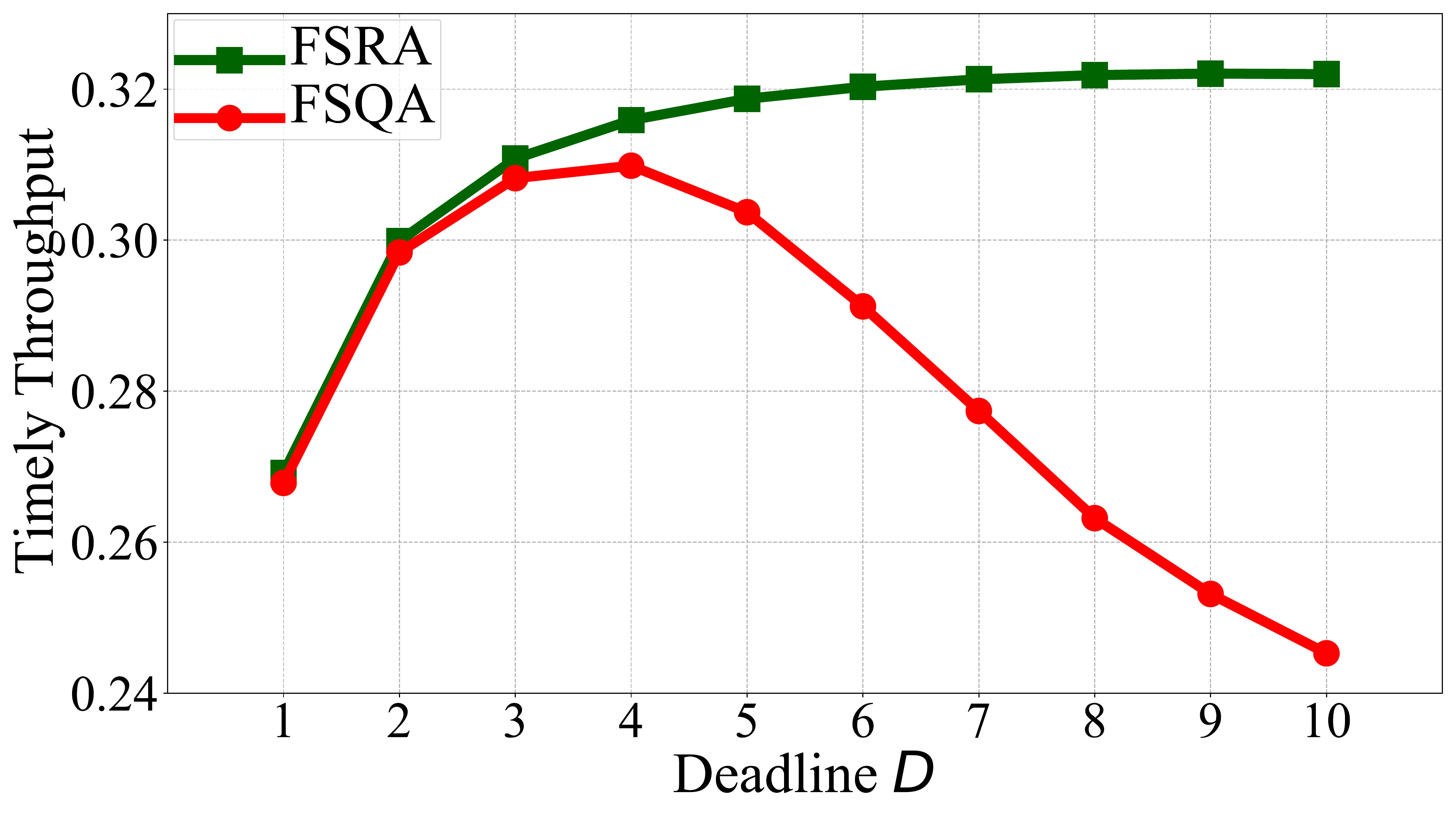}
  \caption{Comparison of the system timely throughputs of FSRA and FSQA for the two-device problem.}\label{fig:compare_rl_ql}
\end{figure}

We compare FSQA and FSRA by varying $D$ from 1 to 10.
For each $D$, we randomly select 500 groups of different system parameters, i.e., ($p_b$, $p_b'$, $p_s$, $p_s'$, $p_t$).
For each group of parameters, we simulate 10,000,000 slots independently for FSQA and FSRA,
and evaluate the system timely throughput for the last 100,000 slots. The result is shown in Fig.~\ref{fig:compare_rl_ql}.
As we can see, FSRA outperforms FSQA for all $D$'s, suggesting that indeed R-learning is more suitable to our problem than Q-learning.
As we explained before, R-learning is used to solve model-free MDPs with average reward, while
Q-learning is used to solve model-free MDPs with discounted reward. Our problem turns out to be exactly
a model-free MDP with average reward. That is the main reason that FSRA outperforms FSQA.
In Appendix~\ref{app:how_to_improve_FSQA}, we further present an example to compare the policies of FSRA and FSQA after convergence
and explicitly show that FSRA is better than FSQA. We also propose a method to tune FSQA so as to improve its performance.

In addition, we remark that FSQA is more difficult to converge than FSRA. As the hard delay $D$ increases,
we should expect that the system timely throughput also increases since packets have longer lifetime and thus are more difficult to expire.
However, as $D$ increases, the state space $\mathcal{S}'$ (of size $2^{D+2}$) also increases exponentially.
As a result, both FSRA and FSQA need more slots to converge. But FSQA is much more sensitive to the state-space explosion.
When $D \ge 4$, FSQA cannot converge in 10,000,0000 slots such that its system timely throughput even decreases as $D$ increases, as shown in Fig.~\ref{fig:compare_rl_ql}. We will explicitly compare the convergence speeds of FSQA and FSRA in Sec.~\ref{subsec:compare-convergence}.

Although FSRA outperforms FSQA in terms of both the achieved system timely throughput and the convergence speed,
we point out that FSRA still converges slowly as $D$ increases. For example, in Fig.~\ref{fig:compare_rl_ql},
we need to run 10,000,000 slots such that FSRA converges.
This problem is even more severe when $D$ is larger since
the state space $\mathcal{S}'$ (of size $2^{D+2}$) increases exponentially with the hard delay $D$. Please refer to Sec.~\ref{subsec:compare-convergence}
to see the slow convergence speed of FSRA.
This disadvantage is not acceptable for dynamic heterogeneous wireless networks, since a small change
of the network could cause the system to take a long time to re-converge.
To address this problem, we further explore the problem structure and significantly reduce the state space.
As we mentioned in Sec.~\ref{sec:upper_bound},
it is optimal to first transmit the HoL packet (the most urgent packet) if there are multiple packets in the Device 2's queue at any slot.
Thus, we can imagine that the HoL packet has the biggest impact on the system performance.
In fact, \cite{zhang2020scheduling} has applied this idea to derive a near-optimal
heuristic scheduling policy only based on the lead time of the HoL packet
for wireless downlink with deadline and retransmission constraints.
We can also design a new R-learning random access algorithm only based on the lead time of the HoL packet.
Namely, the state of Device 2 at slot $t$ becomes
\be
s_t \triangleq (h_{t,2}, o_t), \label{equ:state-HSRA}
\ee
where $h_{t,2}$ is the lead time of the HoL packet of Device 2 at slot $t$ and it is 0 by convention
if Device 2 does not have any packet at slot $t$. The state space is denoted by $\mathcal{S}''$,
which is of size $4(D+1)$. The R-learning based algorithm is the same as FSRA except that
the state space changes from $\mathcal{S}'$ to $\mathcal{S}''$. We call this algorithm
\textbf{H}oL-packet-based \textbf{S}tate-space \textbf{R}-learning random \textbf{A}ccess (HSRA),
which is detailed in Algorithm~\ref{alg:rl}.

We can be even more aggressive by only considering if Device 2 has a packet whose lead time is 1.
A packet with lead time 1 means that it will be expire at the end of the current slot if it cannot be delivered
successfully in the current slot. Thus, such a packet is the most urgent one among all packets
in the system. Therefore, we re-define the system state of Device 2 as
\be
s_t \triangleq (f_{t,2}, o_t), \label{equ:state-TSRA}
\ee
where
\be
f_{t,2} =
\left\{
  \begin{array}{ll}
    1, & \hbox{if Device 2 has a packet whose lead time is 1 at slot $t$;} \\
    0, & \hbox{otherwise.}
  \end{array}
\right.
\label{equ:def-f-t-2}
\ee
The state space is denoted by $\mathcal{S}'''$ whose size is only 8 now.
Since the state space is quite small and even not related to the hard delay $D$,
we call this algorithm \textbf{T}iny \textbf{S}tate-space \textbf{R}-learning Random \textbf{A}ccess (TSRA).
Again, TSRA is the same as FSRA except that the state space changes from $\mathcal{S}'$ to $\mathcal{S}'''$,
which is also detailed in Algorithm~\ref{alg:rl}.
Since the state space of TSRA is quite small, it converges much faster than FSRA, as shown in Sec.~\ref{subsec:compare-convergence} shortly.
We will also show that its performance is close to HSRA and FSRA in Sec.~\ref{sec:simulation}.
Thus, this is the final designed policy for our studied two-device problem in Sec.~\ref{sec:system_model}.
We will also extend TSRA to the general multi-device problem in Sec.~\ref{sec:multi-device}.

\begin{figure*}[t]
  \centering
  \subfigure[]{
    \label{fig:convergence_fsqa} 
        \includegraphics[width=0.3\linewidth]{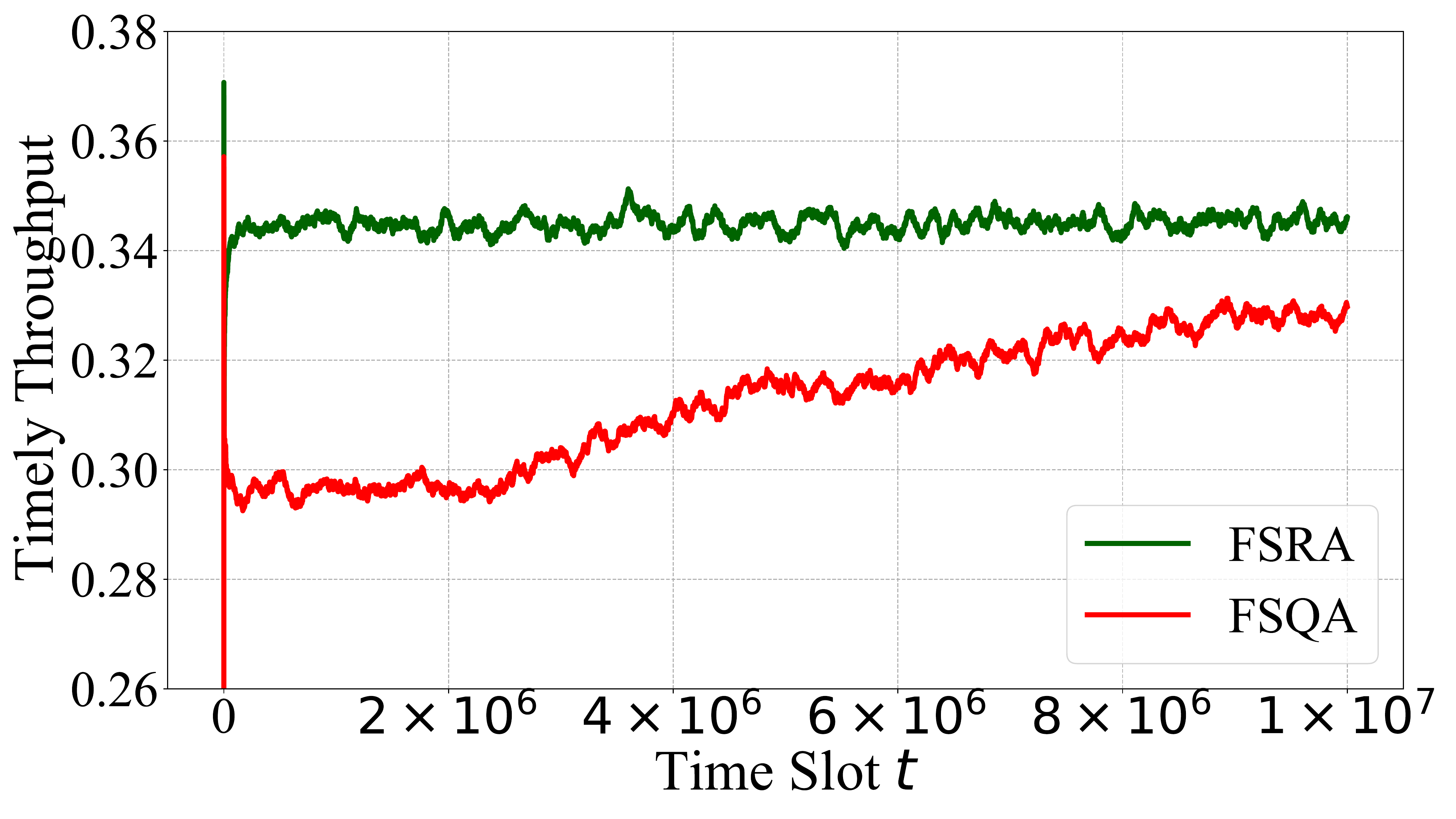}}
    \hfill
    \subfigure[]{
   \label{fig:convergence_fsra}
          \includegraphics[width=0.3\linewidth]{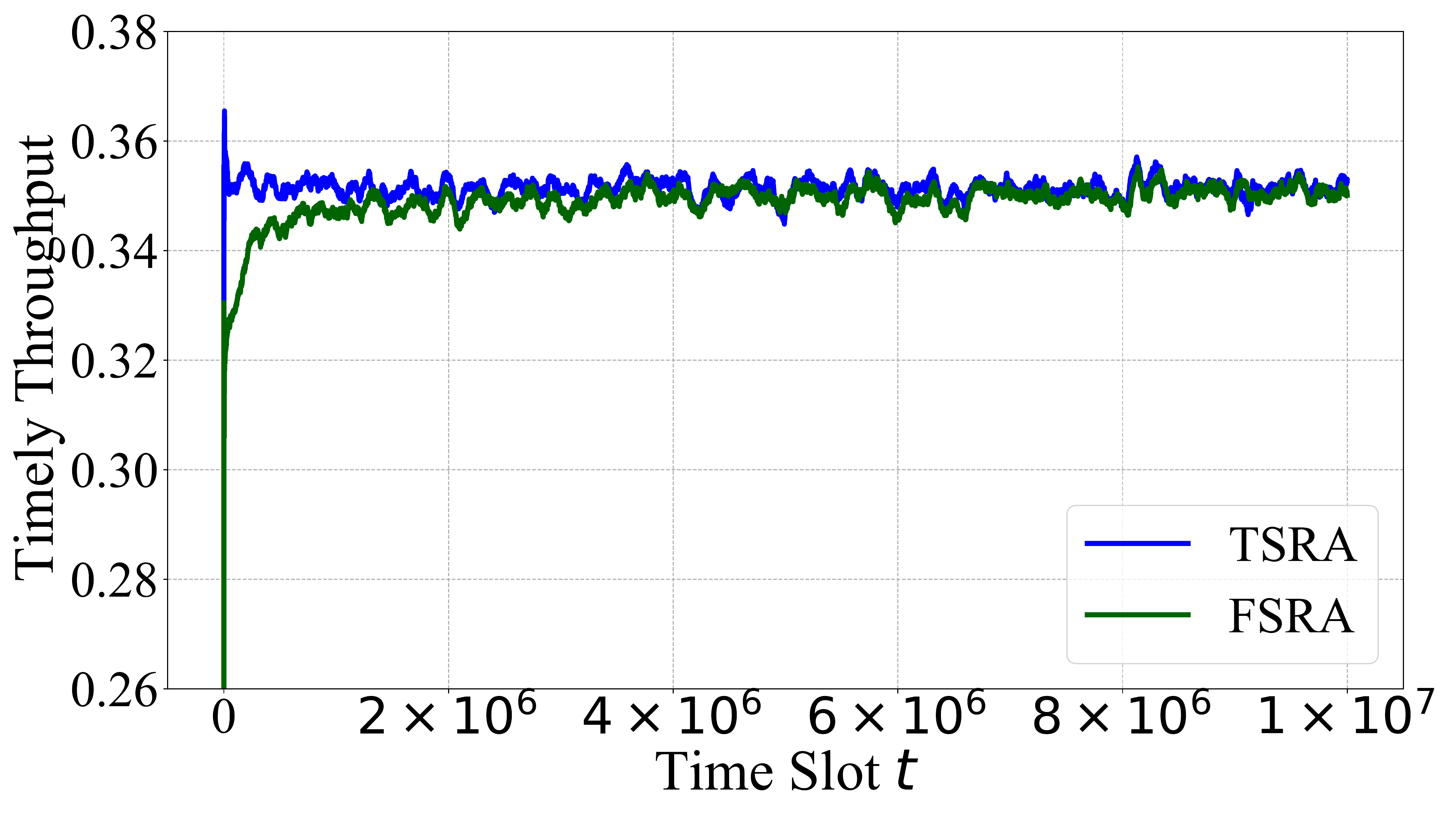}}
    \hfill
    \subfigure[]{
    \label{fig:convergence_tsra}
            \includegraphics[width=0.3\linewidth]{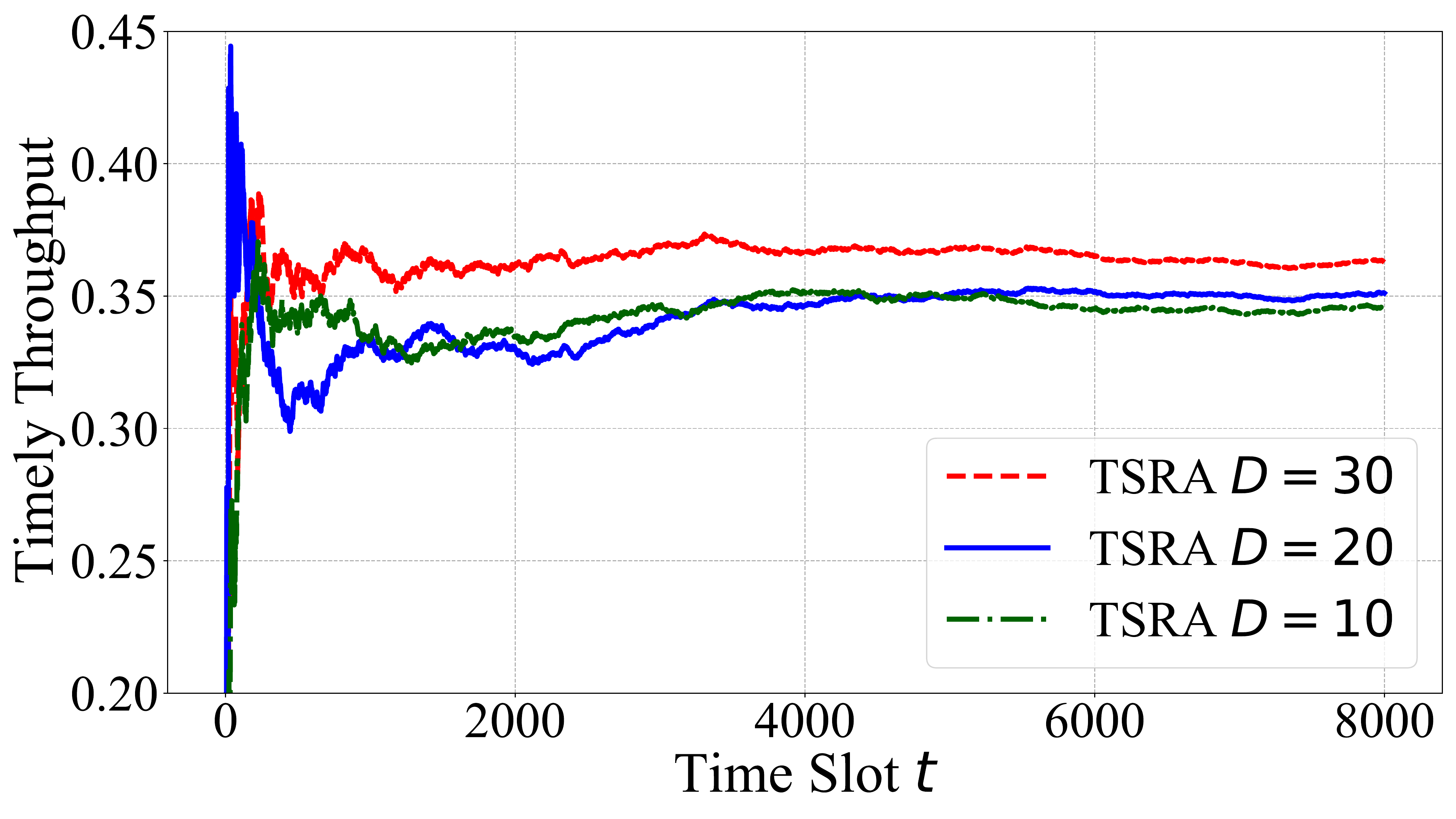} }
  \caption{Comparison of the convergence speeds of different schemes for the two-device problem.
  (a) FSRA v.s. FSQA with $D=5$.
  (b) TSRA v.s. FSRA $D=10$.
  (c) TSRA with three different hard delay values.
\label{fig:convegence-results-two-devices} } 
\end{figure*}

%

\subsection{Comparing Convergence Speeds of FSQA, FSRA and TSRA} \label{subsec:compare-convergence}
In this subsection, we compare the convergence speeds of FSQA, FSRA and TSRA. We set
the system parameters to be $p_b=0.5$, $p_b'=0.4$, $p_s=0.7$, $p_s'=0.6$, $p_t=0.4$,

We first show that FSQA is more difficult to converge than FSRA with $D=5$.
The result is shown in Fig.~\ref{fig:convergence_fsqa}.
We can observe that FSRA converges in 200,000  slots, while FSQA does not converge at the end of the simulation.
Namely, FSQA cannot converge in 10,000,000 slots in this example, which indeed demonstrates that FSQA is very difficult to converge.

We next show that FSRA is more difficult to converge than TSRA  with a larger hard delay $D=10$.
The result is shown in Fig.~\ref{fig:convergence_fsra}.
We can observe that TSRA converges much faster than FSRA, and its achieved system
timely throughput after convergence is almost the same as that of FSRA.

We can further enlarge the hard delay $D$ and show that TSRA still converges very fast.
We let $D$ be 10, 20, and 30, respectively.
The result is shown in Fig.~\ref{fig:convergence_tsra}. We can observe that TSRA converges in 6,000 slots for all three cases.
The fast convergence speed of TSRA makes it suitable in practical dynamic heterogeneous networks.


\begin{figure}[t]
  \centering
  \includegraphics[width=0.6\linewidth]{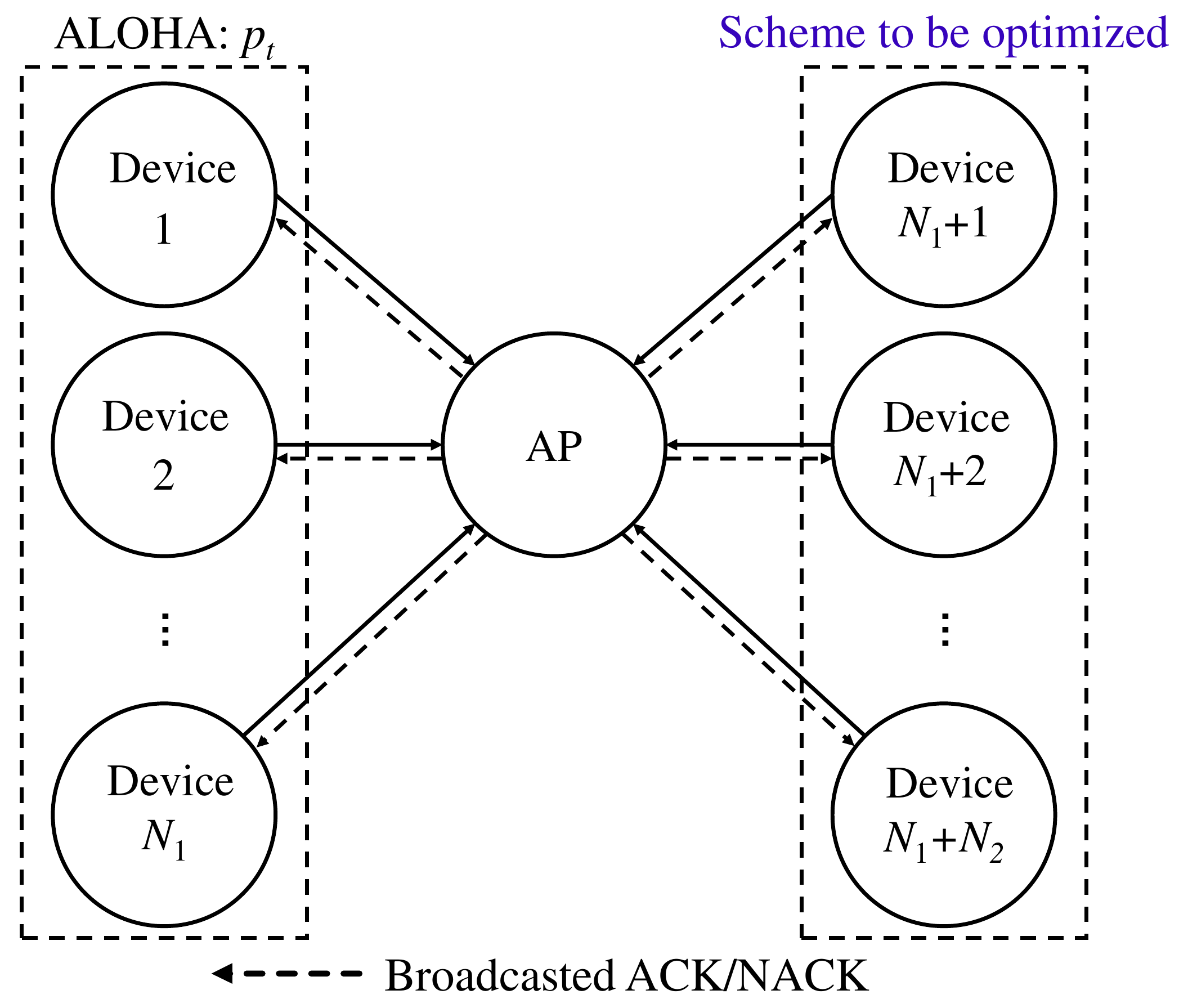}\\
  \caption{The system model for the multi-device case.}\label{fig:model-multiple-device}
\end{figure}

\section{Extension to the Multi-Device Case}  \label{sec:multi-device}
In the previous three sections, we illustrated how and why we design TSRA
for a delay-constrained heterogeneous wireless network
with only two devices. However, the number of devices
in a wireless network could be in the order of hundreds or even thousands.
Thus, we should investigate how to design a random access scheme
for the general multi-device problem. Fortunately, we will soon show
that TSRA after slightly modifying the reward function still works well
for the general multi-device problem.

Similar to the two-device problem, the system model for the multi-device problem is shown in Fig.~\ref{fig:model-multiple-device}.
We assume that there are in total $N$ devices (denoted by set $\mathcal{N}$).
Among them, the first $N_1$ devices (denoted by set $\mathcal{N}_1$) are out of our control,
and we assume that they use ALOHA protocol with the same transmission/retransmission probability $p_t$ for simplicity.
The rest of $N_2=N-N_1$ devices (denoted by set $\mathcal{N}_2 \triangleq \mathcal{N} \setminus \mathcal{N}_1$)  are under our control, and we should design a random access scheme for them.
For example, they could use ALOHA, DLMA, or FSRA/HSRA/TSRA. In this work, we directly apply the TSRA scheme
designed for the two-device problem (see Sec.~\ref{sec:TSRA}) to this multi-device problem. Specifically, similar to \eqref{equ:state-TSRA},
the state of any Device $i \in \mathcal{N}_2$ at slot $t$ is defined as
\be
s_t \triangleq (f_{t,i}, o_t), \label{equ:state-TSRA-multi-device}
\ee
where
\be
f_{t,i} =
\left\{
  \begin{array}{ll}
    1, & \hbox{if Device $i$ has a packet whose lead time is 1 at slot $t$;} \\
    0, & \hbox{otherwise.}
  \end{array}
\right.
\label{equ:def-f-t-i}
\ee
In addition, the action of any Device $i \in \mathcal{N}_2$ at slot $t$ is defined as $a_{t,i} \in \{ \textsf{TRANSMIT}, \textsf{WAIT} \}$.
The TSRA scheme is also the same as that in Algorithm ~\ref{alg:rl}.

\begin{table}[t]
\caption{The new reward function for any Device $i \in \mathcal{N}_2$ for the multi-device problem. \label{tab:new-reward}}
 \scriptsize
 \centering
\begin{tabular}{|c|cc|cc|}
\hline
 \multirow{2}{*}{Observation $o_t$} & \multicolumn{2}{c|}{ $a_{t-1,i} = \textsf{TRANSMIT}$}     & \multicolumn{2}{c|}{$a_{t-1,i} =\textsf{WAIT}$}           \\ \cline{2-5}
                     & \multicolumn{1}{c|}{$f_{t-1,i}=1$}        & $f_{t-1,i}=0$        & \multicolumn{1}{c|}{$f_{t-1,i}=1$}        & $f_{t-1,i}=0$        \\ \hline
\textsf{IDLE}       & \multicolumn{1}{c|}{Impossible} & Impossible & \multicolumn{1}{c|}{-3}         & 2          \\ \hline
\textsf{BUSY}       & \multicolumn{1}{c|}{Impossible} & Impossible & \multicolumn{1}{c|}{10}         & 10         \\ \hline
\textsf{SUCCESSFUL} & \multicolumn{1}{c|}{10}         & 10         & \multicolumn{1}{c|}{Impossible} & Impossible \\ \hline
\textsf{FAILD}      & \multicolumn{1}{c|}{-5}         & -5         & \multicolumn{1}{c|}{2}          & 2          \\ \hline
\end{tabular}
\end{table}

However, we remark that we cannot simply apply the reward function in \eqref{equ:reward function_RL}
to the multi-device case. In \eqref{equ:reward function_RL}, we have only two reward values.
The reward is 1 if the channel observation is \textsf{BUSY} or \textsf{SUCCESSFUL},
while the reward is 0 if the channel observation is \textsf{IDLE} or \textsf{FAILED}.
This two-level reward function works well for the simple two-device problem.
But later in Fig.~\ref{fig:n=10-100-t-o}, we will show  that TSRA with this two-level reward function could have very bad performance.
The main reason is the multi-device problem is much more complicated
than the two-device problem. The multi-device system has many more system possibilities than the two-device problem.
For example, if a device in $\mathcal{N}_2$ observes channel state $\textsf{FAILED}$ and
it does not transmit a packet, in the two-device case, the only reason is that
the other device (i.e., Device 1) transmits a packet and a channel error happens.
However, in the multi-device case, another reason is that more than two other devices simultaneously transmit a packet and thus a channel collision happens. We should use a more sophisticated reward function with more value levels to better differentiate such different system possibilities.
We thus design a new reward function as shown in Table \ref{tab:new-reward}.
Later in Fig.~\ref{fig:n=10-100-t}, we will show that TSRA with this modified multi-level reward function
works much better that the simple two-level reward function in \eqref{equ:reward function_RL}.

In addition, we remark that we cannot control the devices in $\mathcal{N}_1$ who use ALOHA.
At any slot $t$, if at least one device in $\mathcal{N}_1$ transmits a packet,
all devices in $\mathcal{N}_2$ cannot have a successful transmission in this slot.
Instead, such devices in $\mathcal{N}_2$ can even jeopardize the transmission of devices in $\mathcal{N}_1$ if they join the competition.
In other words, If the uncontrollable devices in $\mathcal{N}_1$ are already very congested such that
there is no room for other devices to join the competition, then
there is no significant benefit to add TSRA devices. Therefore, we should focus on the case that
the uncontrollable devices in $\mathcal{N}_1$  are not so congested.
In the extreme case, we should consider that $N_1=0$, i.e., no ALOHA devices but only TSRA devices exist in the system.
This problem is very interesting and important, because it can examine if multiple TSRA devices can
collaborate with each other. Later in Sec.~\ref{sec:simulation}, we will first evaluate the performance of the extreme case
$N_1=0$ and then evaluate the performance of light-loaded cases with $N_1=1,2,3$.

\section{Simulations}\label{sec:simulation}

In this section, we carry out extensive simulations to validate the effectiveness
of our proposed random access scheme TSRA and demonstrate that TSRA outperforms the existing baselines,
including DLMA, which is the random access scheme adopted by \cite{yiding2019deep} for delay-unconstrained heterogeneous wireless networks,
and Learn2MAC, which is an online-learning-based random access scheme proposed in \cite{destounis2019learn2mac} to provide delay guarantee and low energy consumption
for homogeneous wireless networks with frame-synchronized delay-constrained traffic patterns and perfect channel.
We implement all algorithms and evaluate their performances using Python language (7K+ lines of code).
All evaluations are conducted in a computer with two CPUs (Intel Xeon E5-2678 v3), one GPU (NVIDIA GeForce GTX 2080 Ti),
and 64GB memory, running Ubuntu 16.04.6 LTS.
All source codes are publicly available in \textsl{https://github.com/DanzhouWu/TSRA}.

In the following, we will separately show the simulation results for the two-device case and the multi-device case.

\begin{figure}[t]
  \centering
  \includegraphics[width=0.6\linewidth]{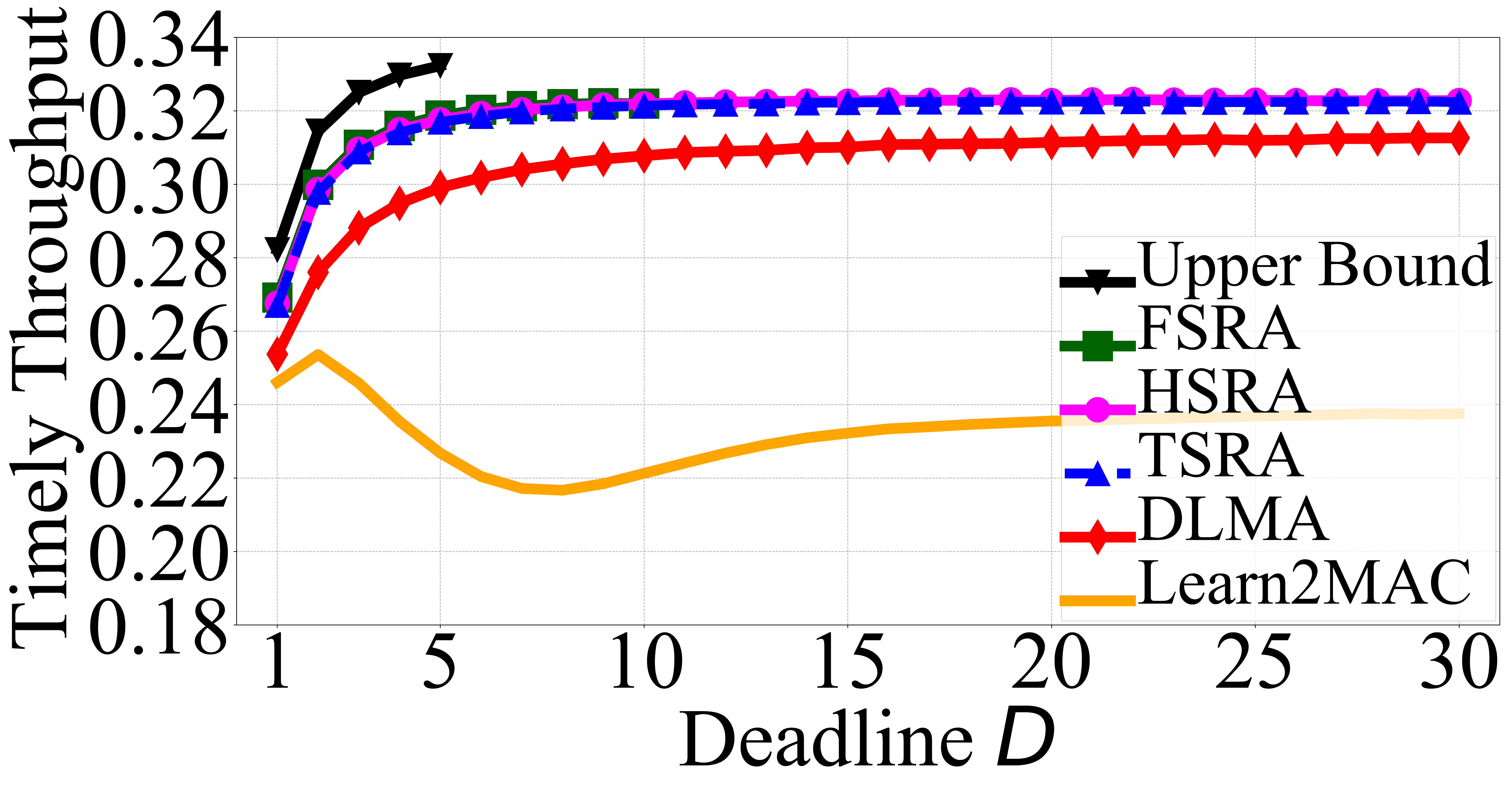}\\
  \caption{Comparison of the system timely throughputs of the upper-bound algorithm, FSRA/HSRA/TSRA, and DLMA for the two-device problem.  \label{subfig:fig:timely_throu}}
\end{figure}

\subsection{Two-Device Case}  \label{subsec:simulation-two}
We first demonstrate the performance of our proposed schemes for the two-device problem.

\textbf{Timely Throughput Comparison.} We first compare all our proposed random access algorithms, including the upper-bound algorithm, i.e., \eqref{equ:upper-bound-policy},
FSRA/HSRA/TSRA proposed in Sec.~\ref{sec:TSRA}, and the existing baselines, DLMA \cite{yiding2019deep} and Learn2MAC\cite{destounis2019learn2mac}.
We simulate the hard delay $D$ from 1 to 30. For each $D$,  we randomly select 500 groups of system parameters ($p_b$, $p_b'$, $p_s$, $p_s'$, $p_t$),
and independently run each group for 10,000,000 slots for FSRA and 100,000 slots  for the other five algorithms.
We then get the average performance of such 500 groups independently for the six algorithms.
The results are shown in Fig.~\ref{subfig:fig:timely_throu}. Note that the state spaces of the upper-bound algorithm and FSRA are of size
$2^{2D+2}$ and $2^{D+2}$, respectively, both of which increase exponentially with $D$.
Due to our computational resource limit, we can only evaluate the upper-bound algorithm for $D \le 5$ and evaluate FSRA for $D \le 10$. Thus, we can see a truncation for both ``Upper Bound" and ``FSRA" curves in  Fig.~\ref{subfig:fig:timely_throu}.

From Fig.~\ref{subfig:fig:timely_throu}, we have the following four observations.
First, the upper bound proposed in Sec.~\ref{sec:upper_bound} indeed provides an effective means for evaluating the timely throughput of different algorithms. This holds by assuming
that Device 2
has more revealed information, including Device 1's parameters and queue information. In addition,
we can numerically quantify the performance gap between the upper bound and any other algorithms.
For example, the system timely throughput of TSRA (resp. DLMA) is 4.98\% (resp. 10.83\%) less than that of the upper bound on
average for $D$ ranging from 1 to 5. Such a performance gap characterization was missing in many other works applying RL to network communication
problems \cite{yiding2019deep,yiding2020non,luong2019applications}. Second, TSRA has very close performance with HSRA and FSRA.
TSRA is only 0.50\% worse than FSRA on average for $D$ ranging from 1 to 10,
and only 0.15\% worse than HSRA on average for $D$ ranging from 1 to 30. This suggests that indeed
we can design the R-learning algorithm only depending on whether Device 2 has a most urgent packet (whose lead time is 1).
Third, our proposed TSRA for delay-constrained heterogeneous wireless networks
achieves better performance than DLMA, which was designed for delay-unconstrained heterogeneous wireless networks.
The system timely throughput of TSRA is 5.62\% larger than that of DLMA on average for $D$ ranging from 1 to 30.
Finally, the Learn2MAC \cite{destounis2019learn2mac} scheme has the worst performance among all schemes. This is because it is designed for
homogeneous network, frame-synchronized traffic pattern and perfect channel, all of which do not hold in our system.

\begin{figure}[t]
  \centering
  \includegraphics[width=0.6\linewidth]{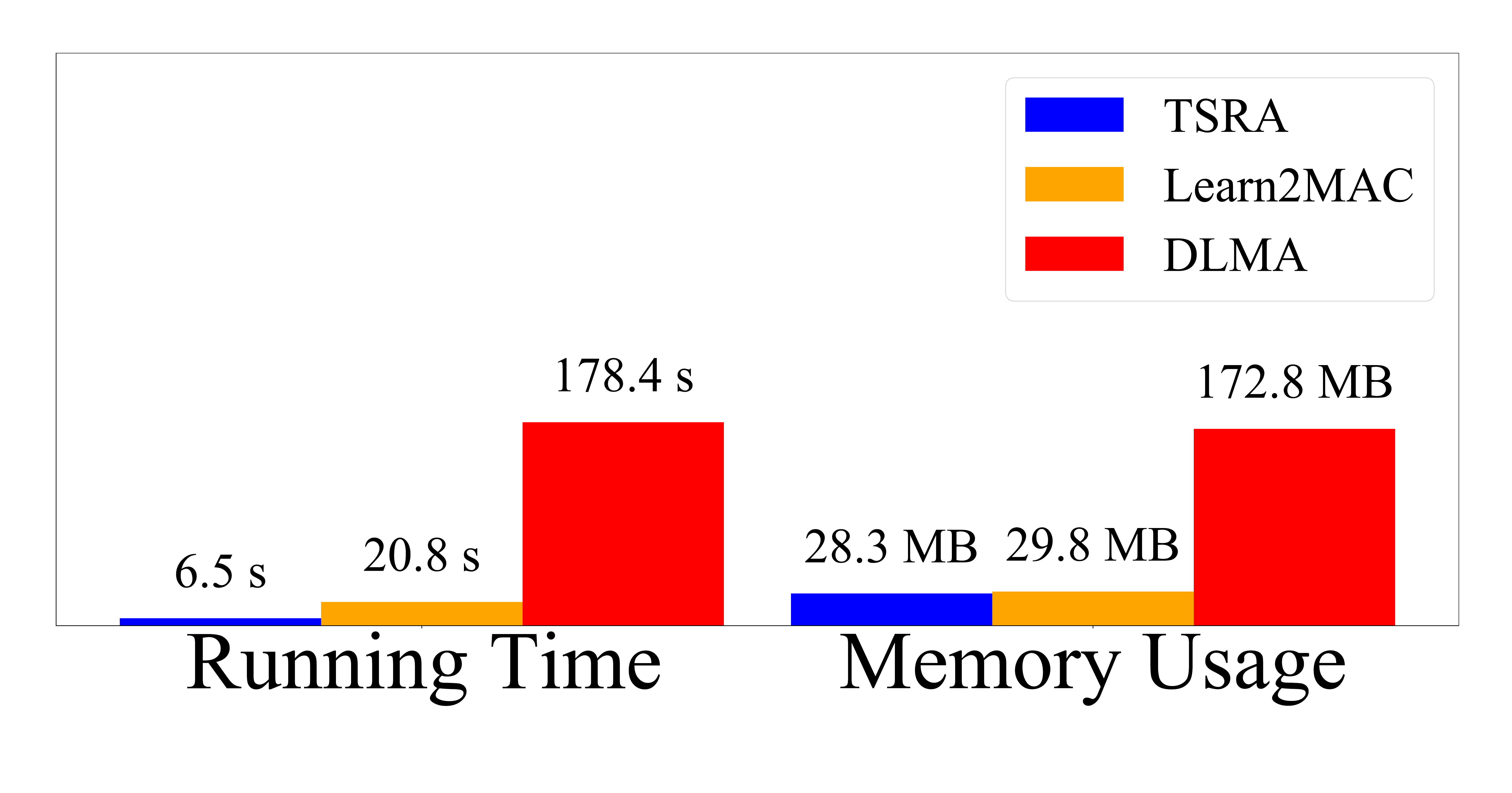}
  \caption{Comparison of the running times and memory usages of TSRA, Learn2MAC (dictionary size = 100), and DLMA with $D=10$ for the two-device problem. \label{subfig:fig:time_memory-two-device} }
\end{figure}

\textbf{Complexity Comparison.} In addition to the performance gain in terms of system timely throughput for TSRA over DLMA and Learn2MAC,
we further use Fig. \ref{subfig:fig:time_memory-two-device}  to demonstrate that TSRA needs significantly less computational resource
than DLMA and Learn2MAC. We run one instance for TSRA, DLMA and Learn2MAC with $p_b=0.5$, $p_b'=0.4$, $p_s=0.7$, $p_s'=0.6$, $p_t=0.4$, and $D=10$.
The total number of running slots is $100,000$ for all three algorithms. As we can see
from Fig. \ref{subfig:fig:time_memory-two-device}, TSRA only needs to run 6.5 seconds,
which is 68.75\% lower than that of Learn2MAC and 96.36\% lower than that of DLMA.
And TSRA only needs 28.3 MB of memory, which is a little bit lower than that of Learn2MAC and is
83.62\% lower than that of DLMA. The reason is as follows.
In terms of time complexity, TSRA only needs to perform two simple computation steps (please refer to \eqref{equ:upgrade_Q} and \eqref{equ:upgrade_rho}) in each slot. Instead, Learn2MAC needs to update the pattern selection probabilities
every $D$ slots and the time complexity depends on the dictionary size (please refer to Step 5 of Algorithm 1 in \cite{destounis2019learn2mac}).
The authors in \cite{destounis2019learn2mac}) recommend the range of the dictionary size to be 100 to 1000, and we choose the conservative value 100 here.
The time complexity could be higher if we choose a larger dictionary size.
DLMA needs to go through a fully-connected multilayer neural network with significantly more computation operations in each slot.
In terms of space complexity, TSRA only needs to store the scalar $\rho$ and the Q-function table $Q(s,a)$,
where $s$ has only 8 possible values and $a$ has only 2 possible values (please refer to Sec.~\ref{sec:TSRA}).
However, DLMA needs to store a memory pool of 500 states, each of which is the collection of transmission events of all past $160$ slots,
and the parameters of the fully-connected multilayer neural network \cite[Table 1]{yiding2019deep}.

\begin{figure}[t]
  \centering
  \subfigure[]{
    \label{fig:robustness-1} 
      \includegraphics[width=0.3\linewidth]{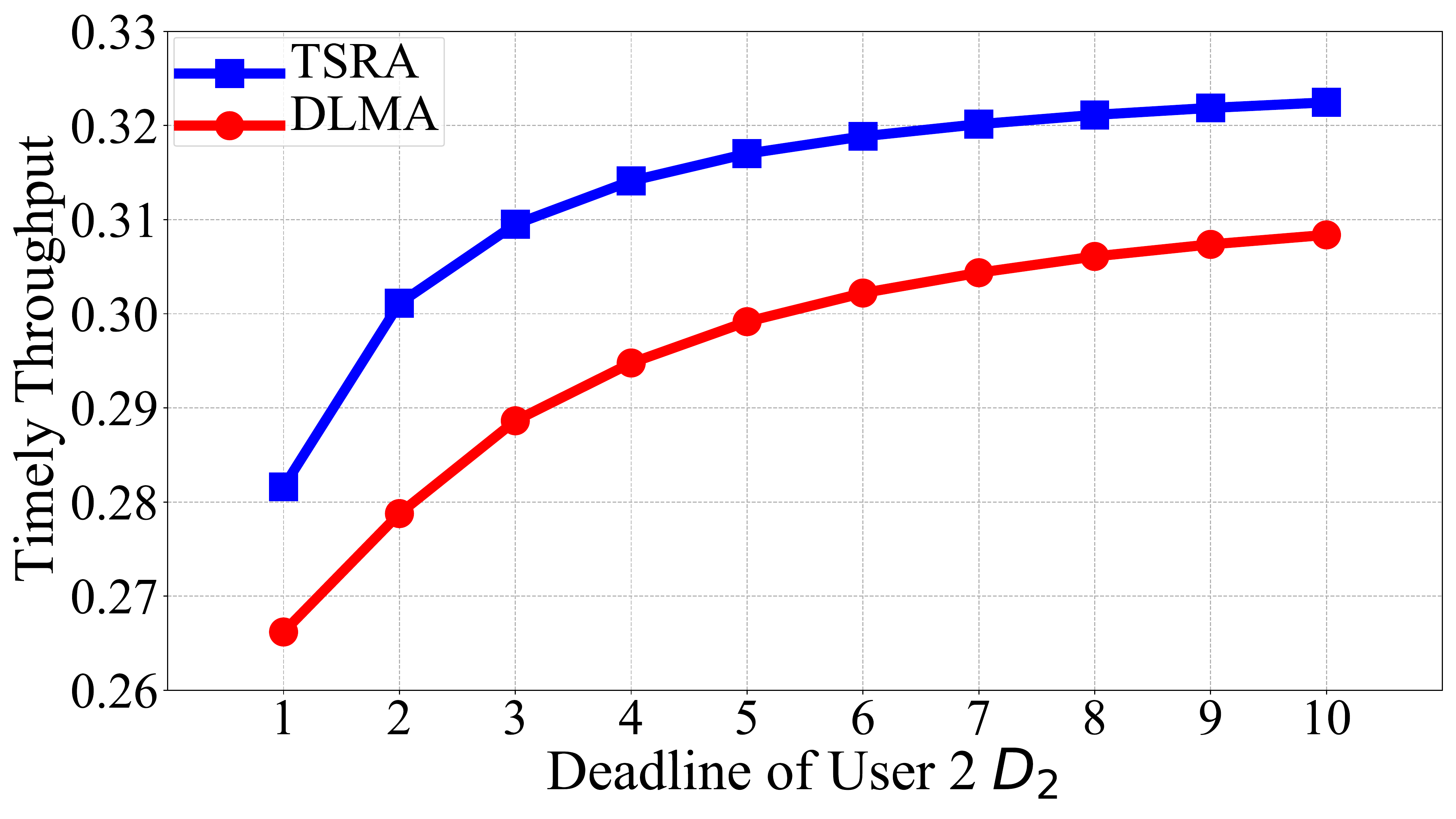}}
    \hfill
    \subfigure[]{
   \label{fig:robustness-2}
          \includegraphics[width=0.3\linewidth]{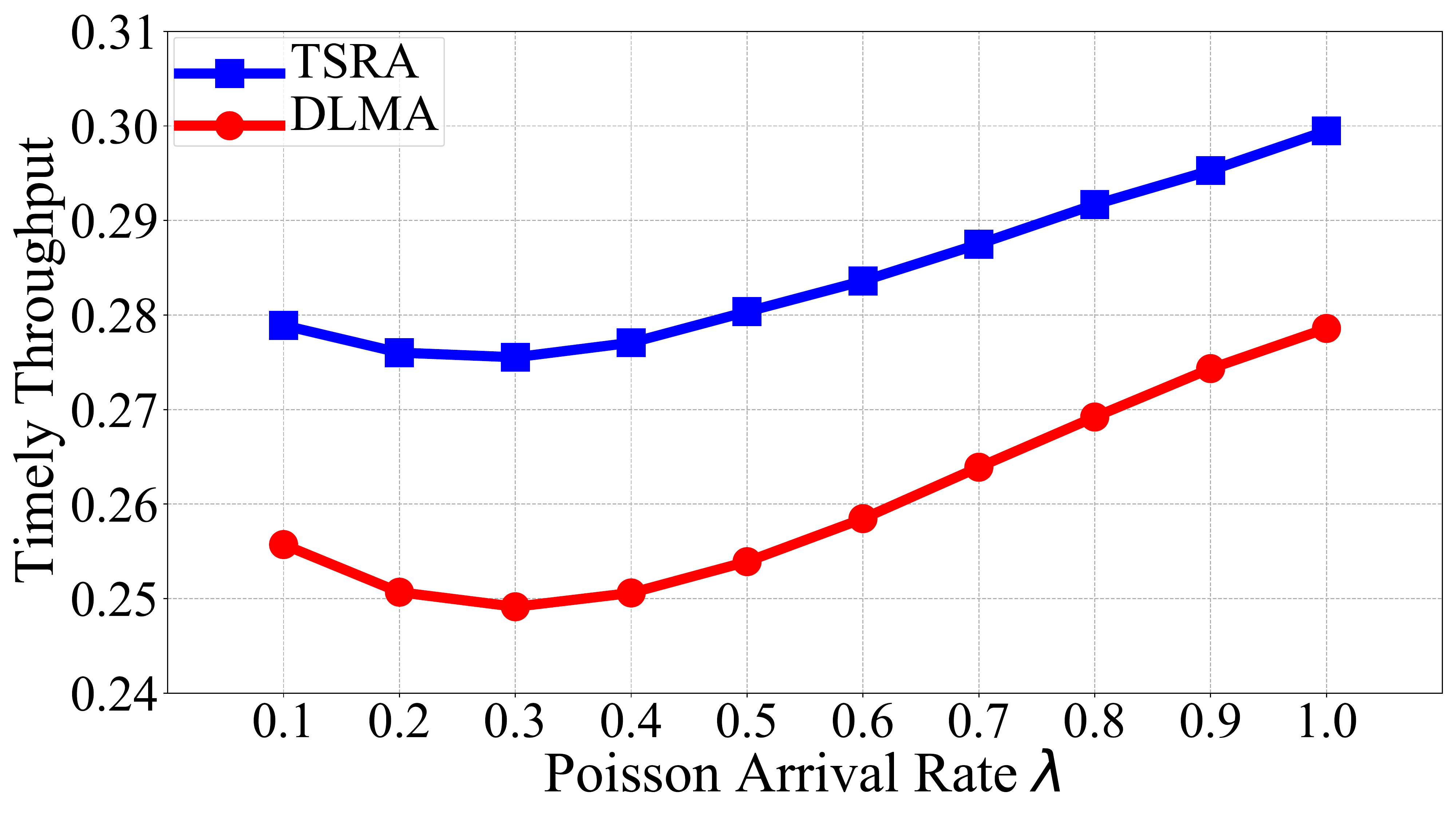}}
    \hfill
    \subfigure[]{
    \label{fig:robustness-3}
            \includegraphics[width=0.3\linewidth]{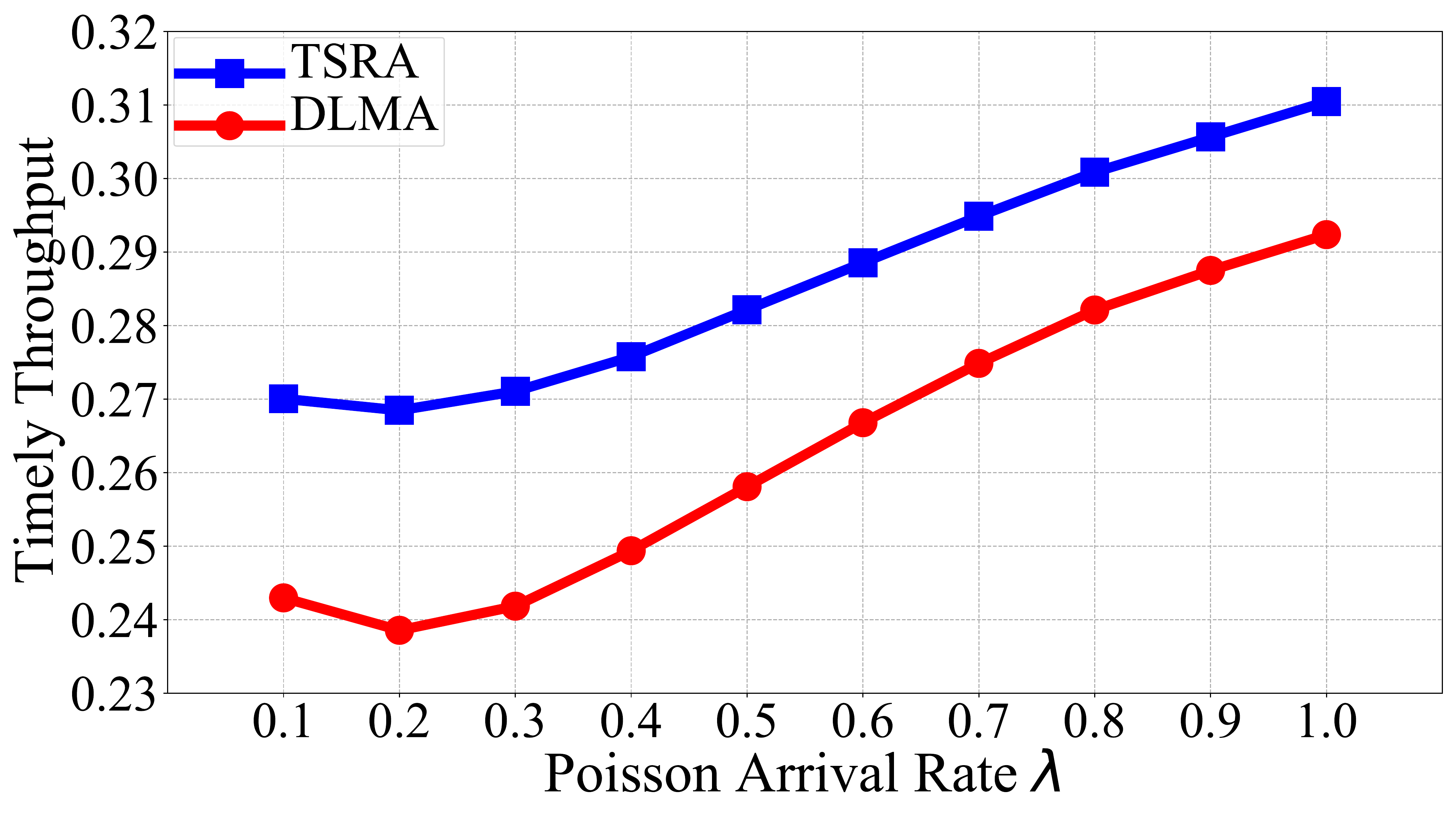} }
  \caption{Robustness results for the two-device problem by comparing the system timely throughputs of TSRA and DLMA.
  (a)  Device 1 and Device 2 have Bernoulli arrivals. We set $D_1=5$ and vary $D_2$ from 1 to 10.
  (b) Device 1 has Poisson arrivals with $D_1=2$ and Device 2 has  Bernoulli arrivals with $D_2=2$. We vary Poisson arrival rate $\lambda$ from 0.1 to 1.0.
  (c) Device 1 has Poisson arrivals with $D_1=4$ and Device 2 has  Bernoulli arrivals with $D_2=2$. We vary Poisson arrival rate $\lambda$ from 0.1 to 1.0.
\label{fig:robustness-results-two-devices} } 
\end{figure}

\textbf{Robustness.} Finally, we demonstrate the robustness of our proposed TSRA algorithm for delay-constrained heterogeneous wireless
 networks.
In this paper, we assume that both devices have Bernoulli arrivals and all their packets have the same hard delay $D$.
We then consider three different settings with larger heterogeneity:
\begin{itemize}
\item Case 1 (Different hard delays): Both Device 1 and Device 2 have Bernoulli arrivals, but they have different hard delay $D$'s  (Fig.~\ref{fig:robustness-1});
\item Case 2 (Different traffic patterns): Device 1 has Poisson arrivals while Device 2 has Bernoulli arrivals, but they have the same hard delay (Fig.~\ref{fig:robustness-2});
\item Case 3 (Different hard delays and different traffic patterns): Device 1 has Poisson arrivals with delay $D_1$, while Device 2 has Bernoulli arrivals with a different delay $D_2$ (Fig.~\ref{fig:robustness-3}).
\end{itemize}

Note that for each point in Figs.~\ref{fig:robustness-1}-\ref{fig:robustness-3},
we get the average among randomly selected 500 groups of system parameters $(p_s,p_t,p'_b,p'_s)$, each running 100,000 slots.
We can observe that TSRA is again better than DLMA for all three cases.
On average, the system timely throughput of TSRA is 5.85\% more than that of DLMA in Fig.~\ref{fig:robustness-1},
9.30\% more that that of DLMA in Fig.~\ref{fig:robustness-2}, and 9.02\% more that that of DLMA in Fig.~\ref{fig:robustness-3}.
These results show that our proposed TSRA is robustly better than DLMA for different heterogeneous settings.

%

\begin{figure*}[t]
  \centering
  \subfigure[]{
    \label{fig:n=10-t} 
        \includegraphics[width=0.48\linewidth]{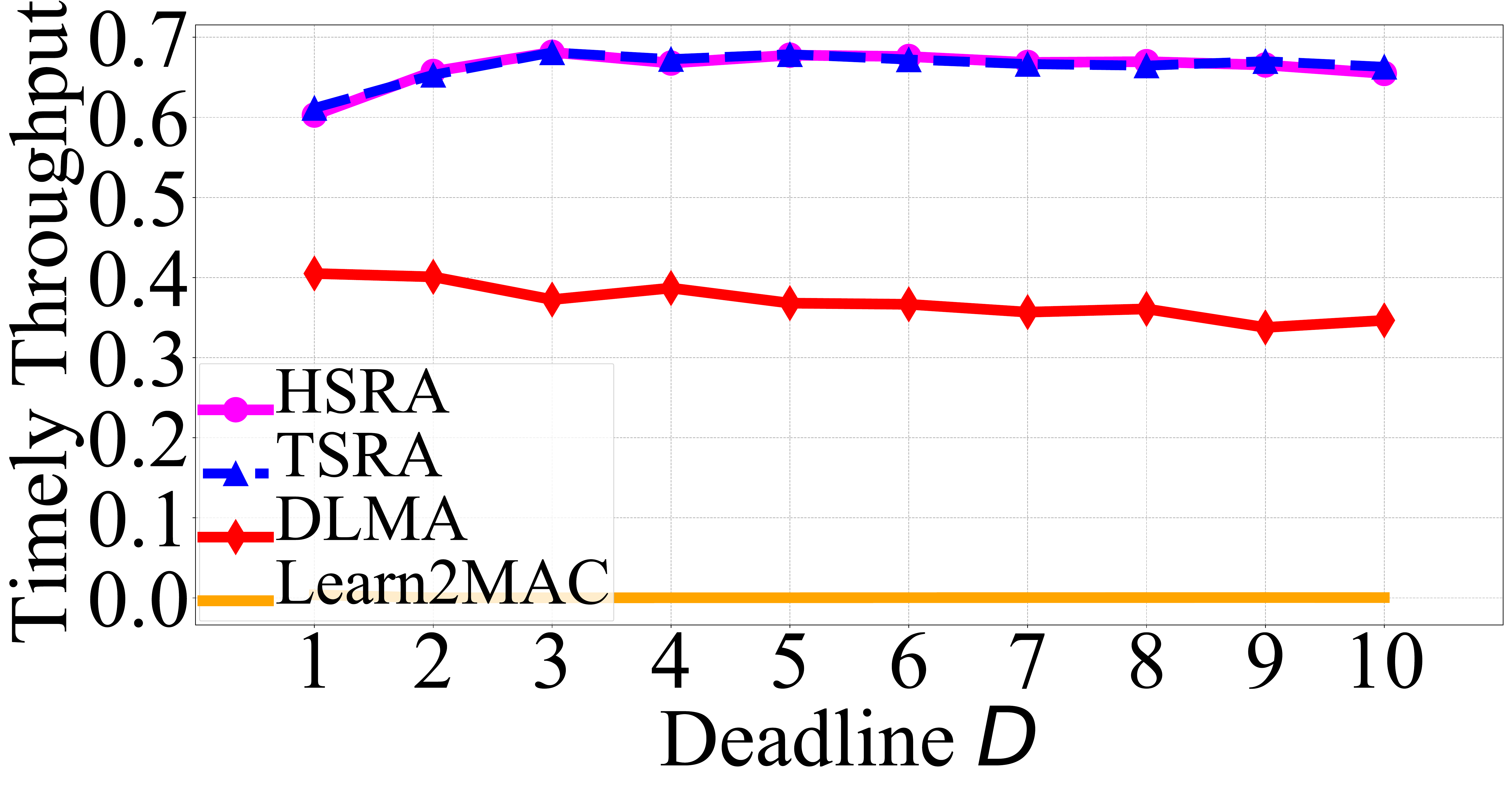}}
    \hfill
    \subfigure[]{
    \label{fig:n=10-100-t-o}
         \includegraphics[width=0.48\linewidth]{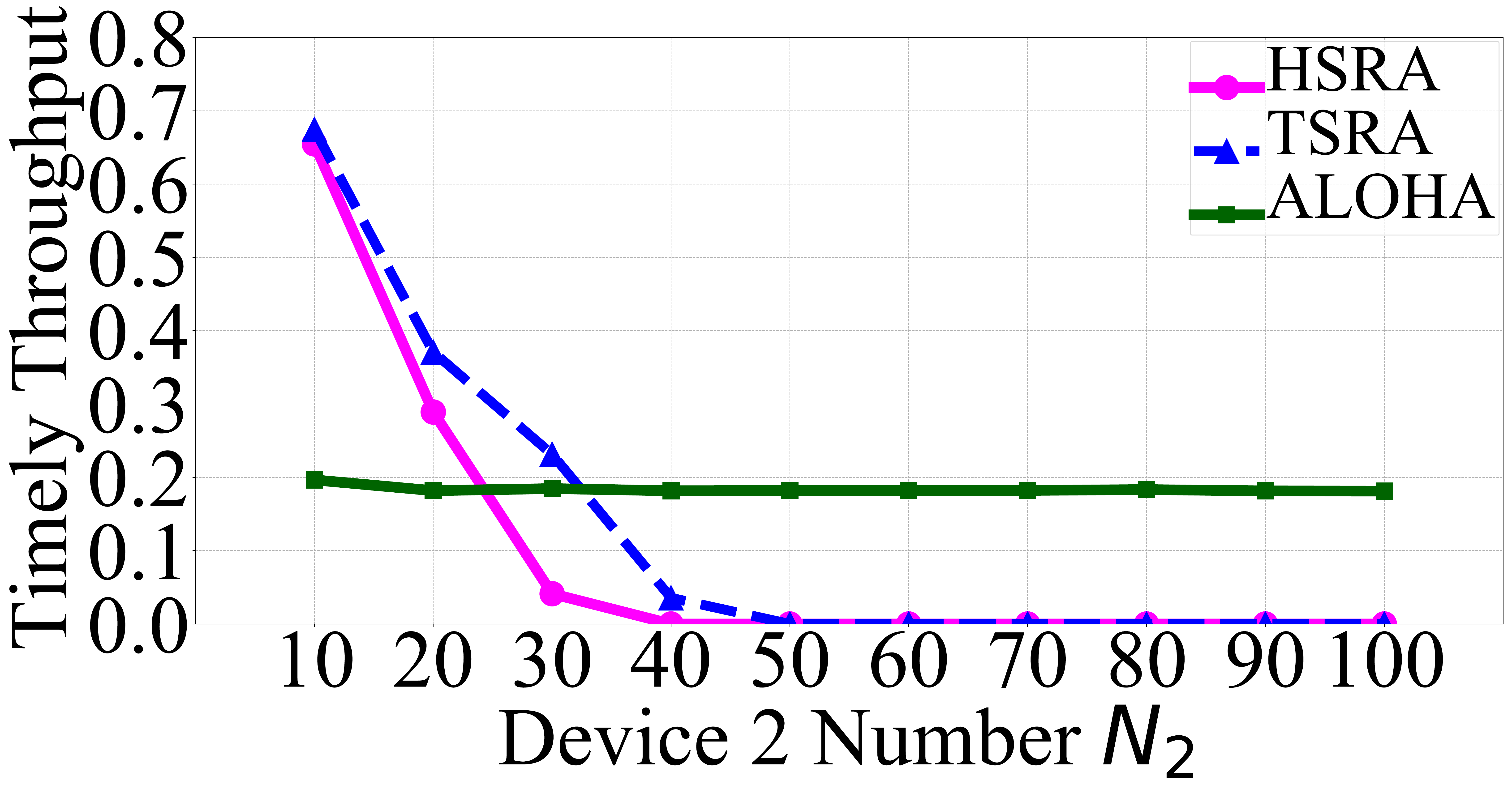}}
    \\
    \subfigure[]{
        \label{fig:n=10-100-t}
      \includegraphics[width=0.48\linewidth]{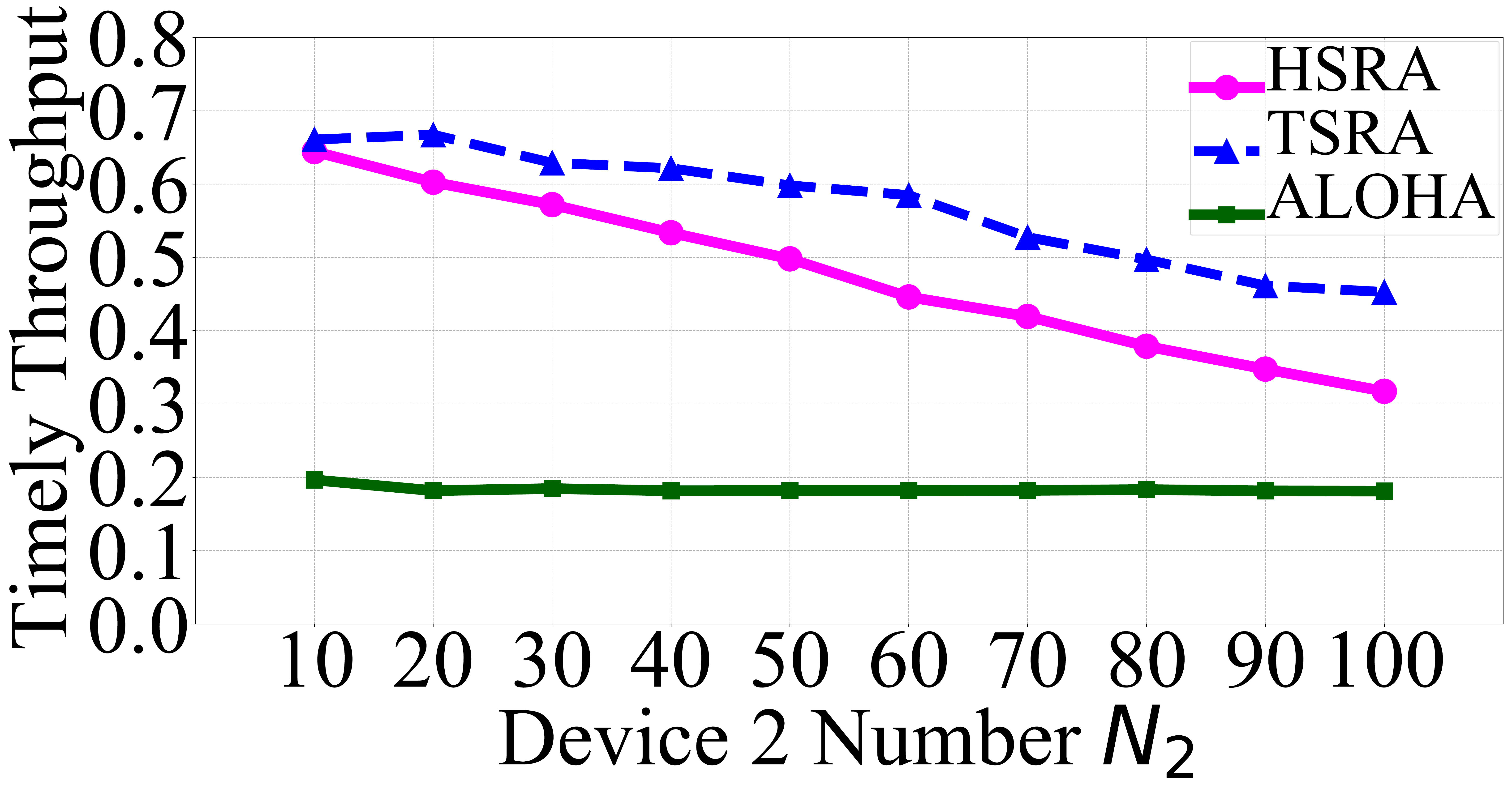}}
  \hfill
  \subfigure[]{
   \label{fig:conv-TSRA-HSRA-multi-device}
        \includegraphics[width=0.48\linewidth]{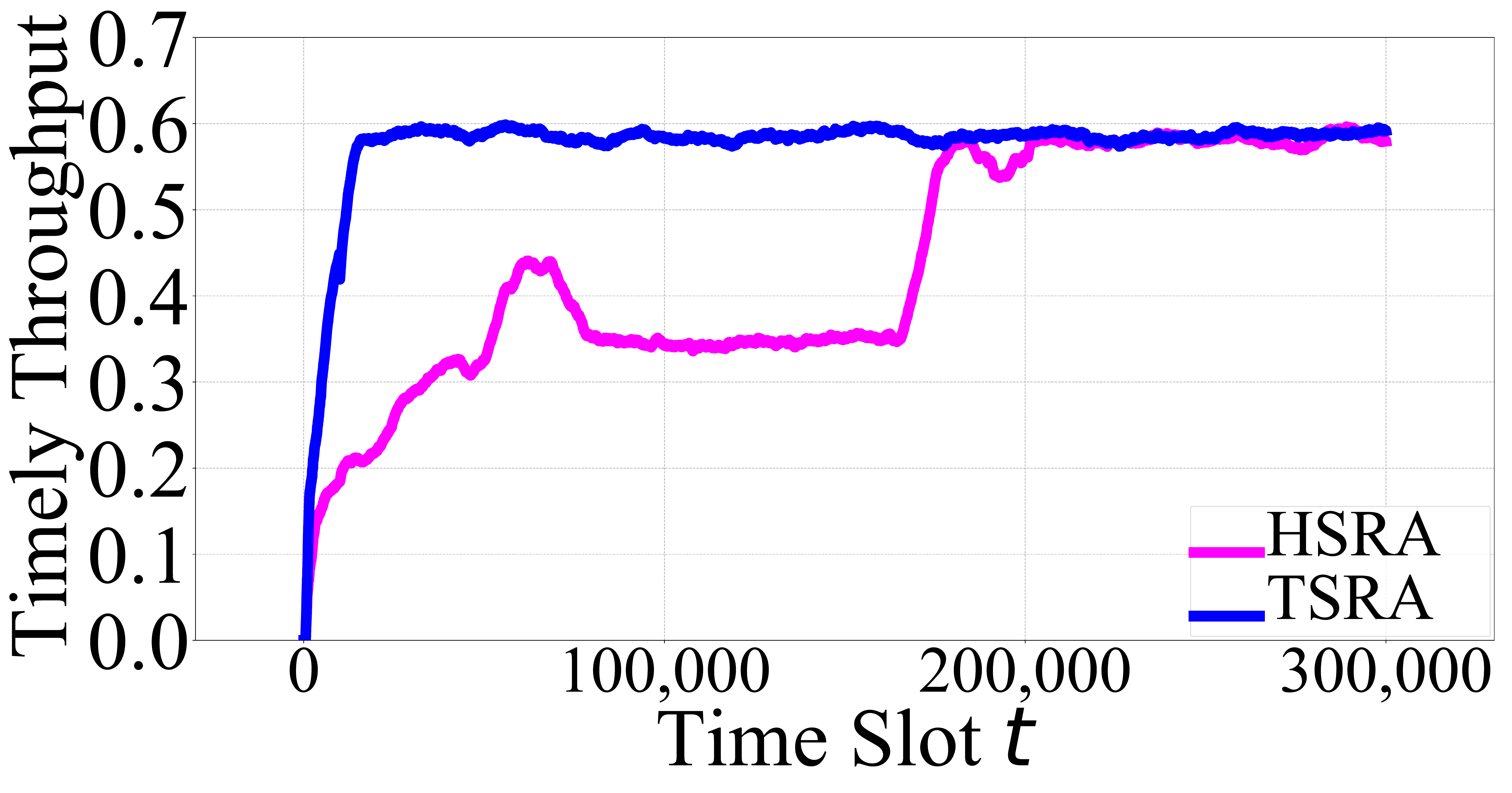}}

  \caption{Comparison of the system timely throughput for the multi-device problem.
  (a) Compare TSRA, HSRA, and Learn2MAC  when $N_1=0$, $N_2=10$ and $D$ from 1 to 10 with reward function in Table \ref{tab:new-reward}.
  (b) Compare TSRA, HSRA, and ALOHA  when $N_1=0$, $D=10$ and $N_2$ varying from 10 to 100 with reward function in \eqref{equ:reward function_RL}.
  (c) Compare TSRA, HSRA, and ALOHA when $N_1=0$, $D=10$ and $N_2$ varying from 10 to 100 with reward function in Table \ref{tab:new-reward}.
  (d) Convergence results of TSRA and HSRA when $N_1=0$, $N_2=100$ and $D=10$ with reward function in Table \ref{tab:new-reward}.
\label{fig:throughput-results-multi-devices} } 
\end{figure*}

\subsection{Multi-Device Case} \label{subsec:simulation-multi}
We now demonstrate the performance of our proposed schemes for the general multi-device problem. In all simulations
in this subsection, we assume that all devices have Bernoulli arrivals.

\textbf{Timely Throughput Comparison.}
We first compare TSRA,  HSRA, DLMA, and Learn2MAC when $N_1=0$ and $N_2=10$.
Namely, in Fig.~\ref{fig:model-multiple-device},
there are zero device in the left side and 10 devices in the right side, all of which will use TSRA/HSRA/DLMA/Learn2MAC.
We vary the hard delay $D$ from 1 to 10. For each $D$,  we randomly select 100 groups of system parameters
and independently run each group for 100,000 slots for all four algorithms.
We then get the average performance of such 100 groups independently for the four algorithms.
The results are shown in Fig.~\ref{fig:n=10-t}. We can see that our proposed TSRA and HSRA achieve the largest
system timely throughput than existing baselines, DLMA and Leanr2MAC.
In addition, TSRA ahd HSRA
have the same performance. This demonstrates that TSRA/HSRA also works well for the general multi-device problem.
Concretely, in this figure, we can see that TSRA increases the system timely throughput by 79.19\% as compared with
DLMA under the multi-device case. The performance gain is significantly larger than that (only 5.62\% in Sec.~\ref{subsec:simulation-two}) of the two-device case.

Now we increase the number of devices, i.e., $N_2$, from 10 to 100 with stepsize of 10 and compare the
two schemes, TSRA and HSRA.  In Sec.~\ref{sec:multi-device}, we have argued that we should change the
reward function from \eqref{equ:reward function_RL} to Table \ref{tab:new-reward}. To justify this modification,
we will independently evaluate TSRA and HSRA for these two different reward functions. As a comparison, we also show the result when all $N_2$ devices use the simple ALOHA scheme where the transmission/retransmission probability is $1/N_2$ \cite{lei2018on}.
Fig.~\ref{fig:n=10-100-t-o} shows the result with the two-level reward function (i.e., \eqref{equ:reward function_RL}),
and Fig.~\ref{fig:n=10-100-t} shows the result with the modified multi-level reward function (i.e., Table \ref{tab:new-reward}).
We can see that both TSRA and HSRA become worse when the number of devices $N_2$ increases and even worse
than the simple ALOHA scheme when $N_2 \ge 30$ under the  the two-level reward function (i.e., \eqref{equ:reward function_RL}).
However, with the modified multi-level reward function (i.e., Table \ref{tab:new-reward}), both TSRA and HSRA work better than ALOHA
in all cases. Fig.~\ref{fig:n=10-100-t-o} and Fig.~\ref{fig:n=10-100-t} justify why we should modify the reward function
for the multi-device problem.

We also obtain a somehow counter-intuitive observation from Fig.~\ref{fig:n=10-100-t}. When $N_2$ is large,
the low-complexity TSRA is even better than the medium-complexity HSRA scheme. As we discussed in Sec.~\ref{subsec:R-learning},
HSRA considers the HoL packets while TSRA only considers the most urgent packets. It seems that TSRA losses more system information
than HSRA. Thus, TSRA should be worse than HSRA, or at least same as HSRA. It cannot be strictly better than HSRA,
as justified by Fig.~\ref{subfig:fig:timely_throu} when $N_1=1$ and $N_2=1$ and Fig.~\ref{fig:n=10-t} when $N_1=0$ and $N_2=10$.
However, Fig.~\ref{fig:n=10-100-t} shows that TSRA is strictly better than HSRA when $N_2 \ge 20$. To explain this counter-intuitive observation,
in Fig.~\ref{fig:conv-TSRA-HSRA-multi-device}, we show the convergence result of TSRA and HSRA when $N_1=0$, $N_2=100$ and  $D=10$ with reward function in Table \ref{tab:new-reward}. We can see that HSRA converges much more slowly than TSRA due to its large state space
in the multi-device case. Since we only run 100,000 slots, HSRA has not yet converged and thus its system timely throughput is lower
than that of the converged TSRA.

Therefore, Fig.~\ref{fig:throughput-results-multi-devices} demonstrates the superior performance of TSRA with
the modified multi-level reward function (i.e., Table \ref{tab:new-reward}) for the general multi-device problem.

\begin{figure}[t]
  \centering
  \includegraphics[width=0.6\linewidth]{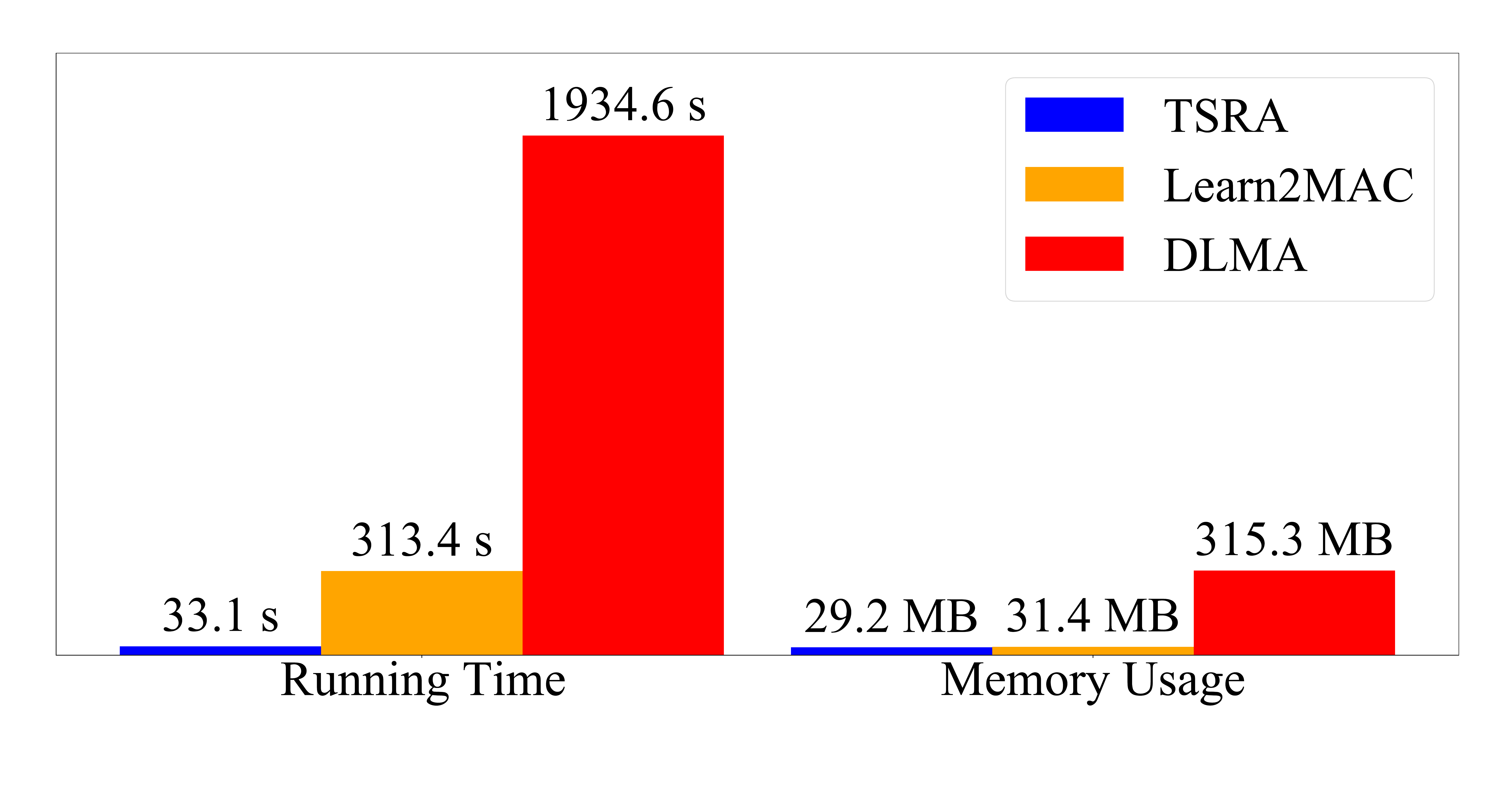}
  \caption{Comparison of the running times and memory usages of TSRA, Learn2MAC (dictionary size = 100), and DLMA when
   $N_1=0$, $N_2=10$ and $D=10$ for the multi-device problem. \label{subfig:fig:time_memory_multi} }
\end{figure}

\textbf{Complexity Comparison.} Similar to Fig.~\ref{subfig:fig:time_memory-two-device}, we also compare the computation
complexity in terms of running time and memory usage for TSRA, DLMA and Learn2MAC for the general multi-device problem.
We consider $N_1=0$, $N_2=10$, and $D=10$. The result is shown in Fig.~\ref{subfig:fig:time_memory_multi}.
We can see that, in terms of running time, TSRA needs to run 33.1 seconds,
which is  89.44\% lower than that of Learn2MAC and  98.29\% lower than that of DLMA.
In terms of memory usage, TSRA  consumes  29.2 MB of memory,
which is 7.01\% lower than that of Learn2MAC and only 90.74\% lower than that of DLMA.
This figure shows that TSRA again has much lower computation complexity than existing baselines
for the general multi-device problem.

\textbf{System Power Consumption Comparison.}
Mobile devices, especially for Internet of Things (IoT) devices, usually have limited battery capabilities and need to work for a long time.
Thus, power consumption is a key performance metric for mobile devices. Although in
the previous sections, we did not optimize the power consumption, in this part, we will
compare the system power consumption of our proposed TSRA scheme and
existing baselines, Learn2MAC and DLMA, for general multi-device problem.
Same as \cite{destounis2019learn2mac}, we define the system power consumption as
the average number of devices that transmit a packet per slot, i.e.,
\be
P = \lim_{T \to \infty} \frac{\sum_{t=1}^{T} \sum_{i \in \mathcal{N}} 1_{ \{\text{Device $i$ transmits a packet at slot $t$}\} }}{T}. 
\label{equ:def-system-power}
\ee
Note that in  \eqref{equ:def-system-power}, we do not model the detailed power consumption in wireless communication, but instead use the average transmission times 
to reflect the power consumption. Since we do not optimize the power allocation in the physical layer,
\eqref{equ:def-system-power} is a simplified but useful performance metric to evaluate the system power consumption in our problem.  

\begin{figure}[t]
  \centering
  \includegraphics[width=0.6\linewidth]{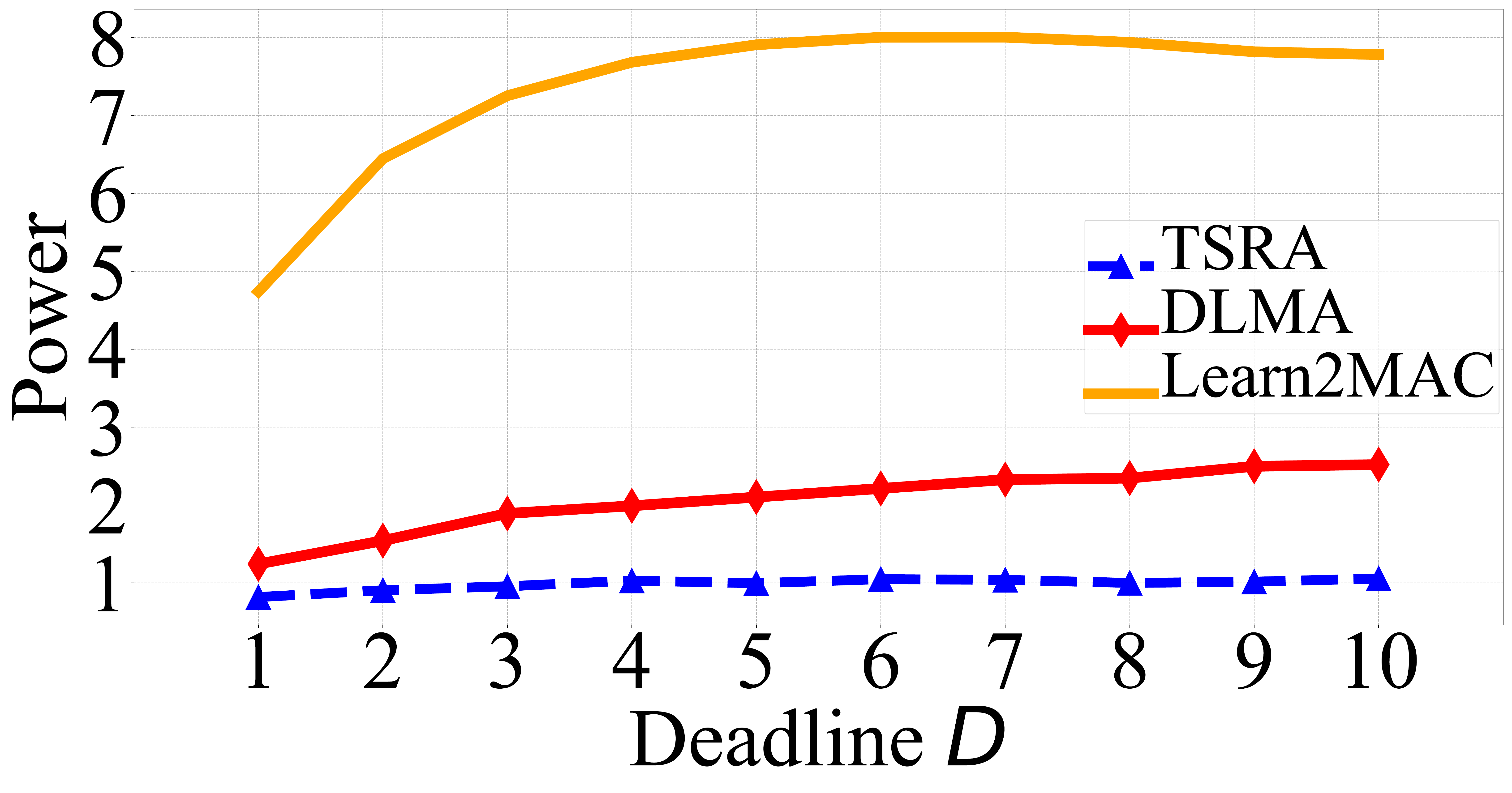}
  \caption{Comparison of power of TSRA, DLMA, and Learn2MAC when
   $N_1=0$, $N_2=10$ and $D$ ranges from 1 to 10 for the multi-device problem. \label{subfig:fig:energy-n1-0} }
\end{figure}

We consider $N_1=0$, $N_2=10$ and $D=10$. The result is shown in Fig.~\ref{subfig:fig:energy-n1-0}.
The system power consumption of TSRA is 0.99 $ \approx 1$, which is 52.29\% lower than that of DLMA and 86.60\% lower than that of Learn2MAC.
This is the best result for random access: if there is one and only one device transmits a packet in each slot,
the system can fully utilize the channel without causing collision.
Thus, TSRA can best collaborate with each other to fully utilize the channel in a distributed manner.
On the contrary, the system power consumption of Learn2MAC is 7.36, implying that on average 7.36 devices
transmit a packet per slot. Thus, collision happens in almost all slots.
This also explains why the system timely throughput of Learn2MAC is almost 0 in Fig.~\ref{fig:n=10-t}.
The system power consumption of DLMA is significantly lower than Learn2MAC, but sill larger than twice of our proposed TSRA scheme.

\textbf{Effect of Different Uncontrollable Congestion Levels.}
As we discussed in Sec.~\ref{sec:multi-device}, we first considered the important case of $N_1=0$.
Thus, all previous simulations in this subsection is based on $N_1=0$.
But it is also very important to see the performance of TSRA when $N_1>0$. Recall that
$N_1$ is the number of uncontrollable ALOHA devices (see Fig.~\ref{fig:model-multiple-device}) in our considered heterogenous wireless network.
In this part, we demonstrate the effect of different
uncontrollable congestion levels by considering $N_1=1,2$ and 3.

\begin{figure}[t]
  \centering
  \subfigure[]{
    \label{fig:congestion-1} 
      \includegraphics[width=0.3\linewidth]{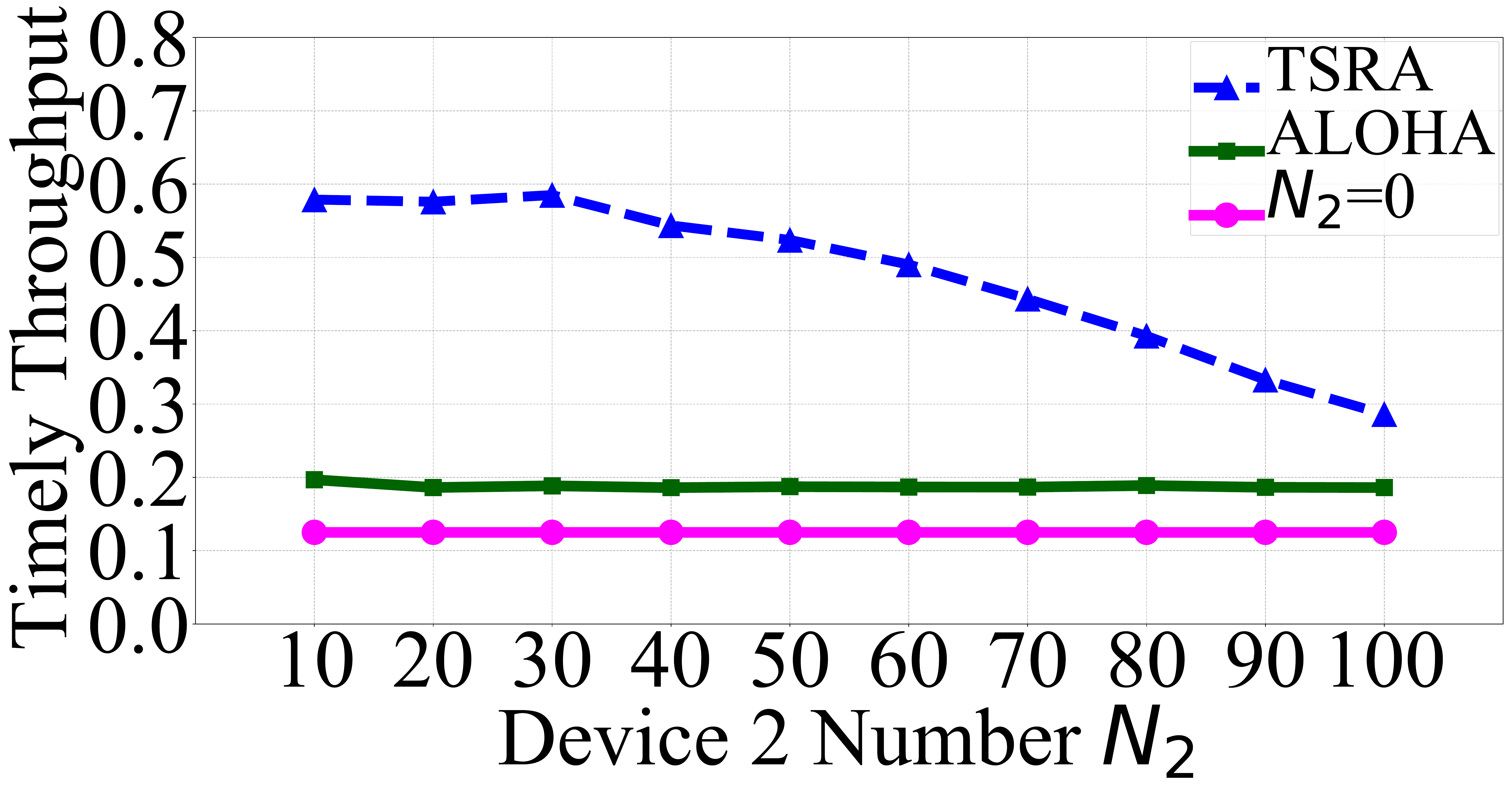}}
    \hfill
    \subfigure[]{
   \label{fig:congestion-2}
          \includegraphics[width=0.3\linewidth]{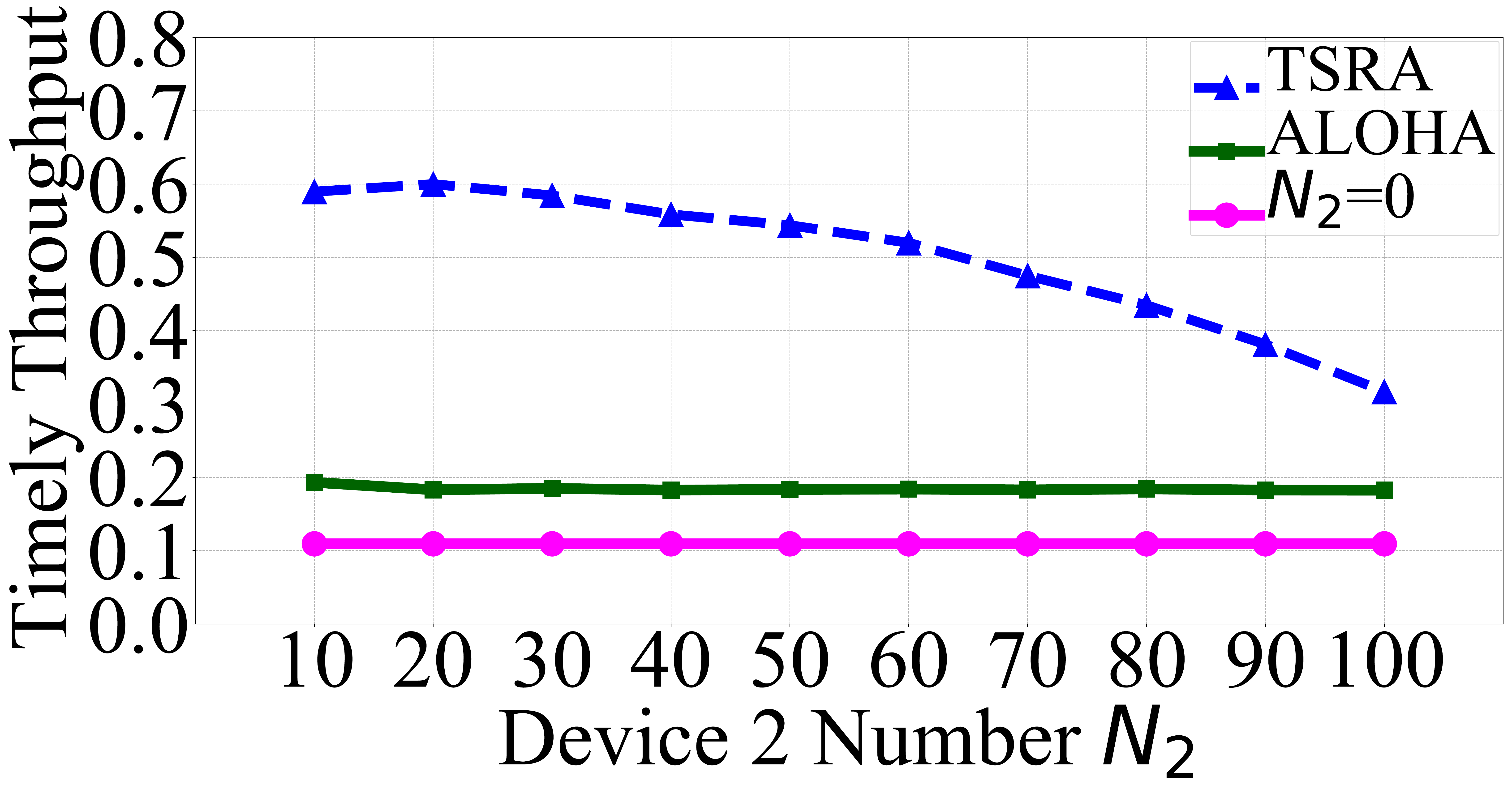}}
    \hfill
    \subfigure[]{
    \label{fig:congestion-3}
            \includegraphics[width=0.3\linewidth]{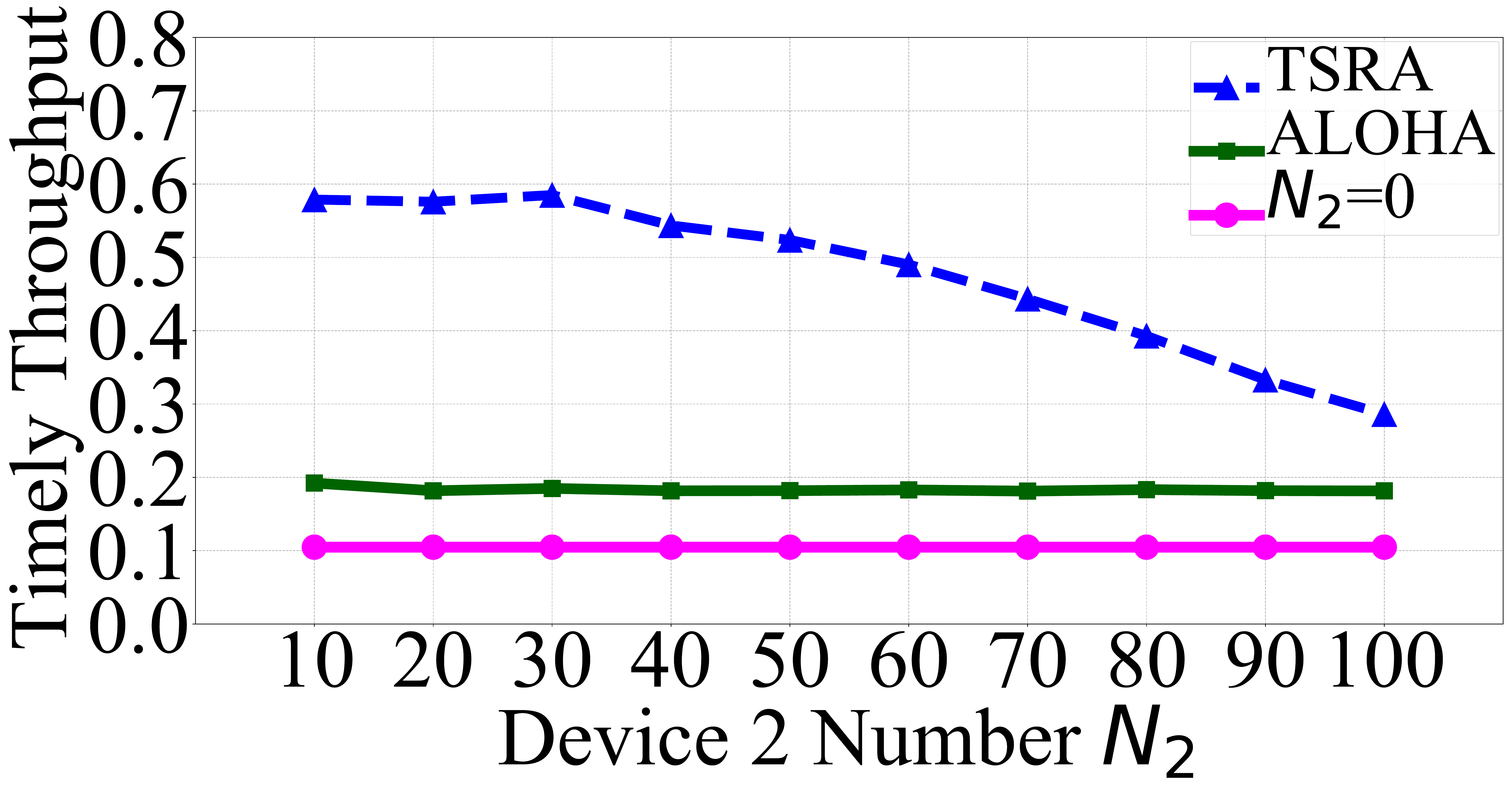} }
  \caption{ Comparison of system timely throughput of TSRA and ALOHA when
  $N_2 \in \{0\} \cup [10, 100]$ and $D=10$ with different $N_1$ values (i.e., different uncontrollable congestion levels)
  for the multi-device problem. Note that $N_2=0$ means that there is zero device in the right side of Fig.~\ref{fig:model-multiple-device}, which is the magenta curve in the figure.
  (a) $N_1=1$.
  (b) $N_1=2$.
  (c) $N_1=3$.
\label{fig:congestion-results} } 
\end{figure}

We assume that all $N_1$ uncontrollable devices has a new packet arrival every slot (i.e., the arrival probability is 1 under the Bernoulli traffic model). In addition, the transmission probabilities of them
are set as $\frac{1}{4N_1}$ and the channel success probabilities are set as 0.5.
We independently simulate both $N_2=0$ and $N_2>0$. Note that $N_2=0$ means that
there are only $N_1$ uncontrollable devices in the system. It can provide a benchmark
to gauge the performance change of adding extra TSRA/ALOHA devices (i.e., $N_2>0$).
For $N_2>0$, we assume that such $N_2$ controllable devices deploy TSRA or ALOHA (with transmission probability $\frac{1}{N_2}$).
For each $N_1$ and $N_2$, we randomly select 100 groups of system parameters and independently run each group for 100,000 slots for TSRA and ALOHA.
We then get the average performance of such 100 groups independently for TSRA and ALOHA.
The result is shown in Fig.~\ref{fig:congestion-results}.
We remark that we cannot run DLMA in our computer platform as DLMA has much higher time and space complexity (see Fig.~\ref{subfig:fig:time_memory_multi}).
In addition, the performance of Learn2MAC is very bad with almost zero system timely throughput (see Fig.~\ref{fig:n=10-t}). Thus, we do not
have DLMA and Learn2MAC results in Fig.~\ref{fig:congestion-results}.

We can see that in all cases, TSRA achieves much higher system timely throughput (on average 161.16\% among all cases) than the simple ALOHA scheme.
In addition, comparing with the case $N_2=0$ (the magenta curve in Fig.~\ref{fig:congestion-results}),
adding extra $N_2$ ALOHA devices can increase the system timely throughput by 63.85\% on average among all cases,
while adding extra $N_2$ TSRA devices can significantly increase the system timely throughput by 327.90\% on average among all cases.
Therefore, our proposed TSRA allows  controllable devices to smartly collaborate with uncontrollable devices
and can opportunistically utilize the channel to further improve the system timely throughput.

\section{Conclusion}\label{sec:conclusion}

In this paper, we for the first time investigate the random access problem for delay-constrained heterogeneous wireless networks.
We propose an R-learning-based low-complexity algorithm, called TSRA.
We show that TSRA achieves higher timely throughput, lower computation complexity, and lower power consumption simultaneously than the existing baseline DLMA \cite{yiding2019deep},
which was designed for delayed-unconstrained heterogeneous wireless networks.

Three key messages have been delivered by this work.
First, although RL has been widely used in many network decision problems,
few works characterize their performance gap  due to  RL's black-box nature.
In this work, for the simple two-device case, we instead propose an MDP-based formulation to derive a model-based upper bound
such that it can numerically quantify the performance gap of any RL-based scheme. We believe that this methodology can
benefit other network problems utilizing RL. Second, since network problems are concerned with throughput or timely throughput, which is by nature an average reward, it is revealed by this work that
average-reward-based R-learning is better than the currently widely-used discounted-reward-based Q-learning.
Finally, for delay-constrained communications,
we show that the most urgent packets have the biggest impact on the system performance,
which can be utilized to simplify the system design significantly.

For future research of this ongoing work, it is interesting and important to consider the device fairness, whilst maximizing the system timely throughput.


%
%

\bibliographystyle{IEEEtran}
\bibliography{ref}

\newpage

\begin{appendix}

\subsection{Proof of Theorem~\ref{thm:D=1}} \label{app:proof-of-theorem-D=1}
For $D=1$, as we explained in Sec.~\ref{sec:upper_bound}, all slots are decoupled such
that we only need to focus on one particular slot. Thus,
we can shrink our design space to a single parameter, i.e., the transmission probability of Device 2, which is denoted by $p'_t$.
To optimize $p'_t$, we first define the following random variables:
\begin{itemize}
\item Random variable $A_i, i=1,2$:
\be
A_i=
\left\{
     \begin{array}{ll}
       1, & \hbox{Device $i$ has a non-expired packet to transmit;} \\
       0, & \hbox{otherwise.}
     \end{array}
   \right.
\ee
\item Random variable $X_i, i=1,2$:
\be
X_i=
\left\{
     \begin{array}{ll}
       1, & \hbox{Device $i$ transmits a packet;} \\
       0, & \hbox{otherwise.}
     \end{array}
   \right.
\ee
\item Random variable $Y_i, i=1,2$:
\be
Y_i=
\left\{
     \begin{array}{ll}
       1, & \hbox{Device $i$ transmits a packet successfully;} \\
       0, & \hbox{otherwise.}
     \end{array}
   \right.
\ee
\end{itemize}

Then, given parameters $p_b, p'_b, p_s, p'_s, p_t, p'_t$, we can derive the
distributions for the above random variables. The system timely throughput is,
\bee
R & = \mathbb{E} [Y_1 + Y_2] = \mathbb{E}[Y_1] + \mathbb{E}[Y_2] = P(Y_1=1) + P(Y_2=1). \label{equ:P-Y1=1+Y2=1}
\eee

Next we compute $P(Y_1=1)$ as follows,
\bee
 P(Y_1=1)
& = \sum_{x_1 \in \{0,1\}} \sum_{x_2 \in \{0,1\}} P(Y_1=1|X_1=x_1, X_2=x_2) P(X_1=x_1, X_2=x_2) \nnb \\
&=P(Y_1=1|X_1=1, X_2=0)P(X_1=1, X_2=0)   \label{equ:Y1=1-e1} \\
&= p_s \cdot P(X_1=1, X_2=0) \label{equ:Y1=1-e2} \\
& = p_s \cdot P(X_1=1) \cdot P(X_2=0) \label{equ:Y1=1-e3},
\eee
where \eqref{equ:Y1=1-e1} holds because Device 1 can deliver a packet successfully only if
Device 1 transmits a packet and Device 2 does not transmits a packet in the considered slot,
and \eqref{equ:Y1=1-e2} holds because the transmission events of both devices are independent.

Now let us again use the law of total probability to  compute $P(X_1=1)$ and $P(X_2=0)$ in \eqref{equ:Y1=1-e3},
\bee
 P(X_1=1) & =   \sum_{a_1 \in \{0,1\}} P(X_1=1|A_1=a_1)P(A_1=a_1)  \nnb \\
& = P(X_1=1|A_1=1) P(A_1=1) \nnb \\
& = p_t p_b,  \label{equ:P-X1=1}
\eee
\bee
& P(X_2=0) =  \sum_{a_2 \in \{0,1\}} P(X_2=0|A_2=a_2)P(A_2=a_2) \nnb \\
& = P(X_2=0|A_2=0) P(A_2=0) + P(X_2=0|A_2=1) P(A_2=1) \nnb \\
& = 1 \cdot (1-p'_b) + (1-p'_t) \cdot p'_b \nnb \\
&  = 1 - p'_bp'_t. \label{equ:P-X2=0}
\eee

Inserting \eqref{equ:P-X1=1} and \eqref{equ:P-X2=0} into \eqref{equ:Y1=1-e3}, we obtain that
\be
P(Y_1=1) = p_sp_tp_b(1-p'_tp'_b). \label{equ:P-Y1=1}
\ee

Similarly, we can obtain
\be
P(Y_2=1) = p'_sp'_tp'_b(1-p_tp_b). \label{equ:P-Y2=1}
\ee

Inserting \eqref{equ:P-Y1=1} and \eqref{equ:P-Y2=1} into \eqref{equ:P-Y1=1+Y2=1}, we obtain the system
timely throughput as,
\bee
R &= p_sp_tp_b(1-p'_tp'_b) + p'_sp'_tp'_b(1-p_tp_b) \nnb \\
& = \left[ p'_{s}-(p_{s}+p'_{s})p_{t}p_{b} \right] p'_{t}p'_{b}+p_{s}p_{t}p_{b}. \nnb
\eee
Thus, if
\be
p_t p_b < \frac{p'_s}{p_s + p'_s}, \label{equ:binary-condition}
\ee
the optimal $p'_t$ to maximize the system timely throughput $R$ is
\be
p'_t=1,
\ee
i.e., Device 2 will always transmit its packet if it has one packet.
Otherwise, if \eqref{equ:binary-condition} does not hold,
the optimal $p'_t$ to maximize the system timely throughput $R$ is
\be
p'_t=0,
\ee
i.e., Device 2 will never transmit its packet.
This completes the proof.


\subsection{Why is FSQA worse than FSRA and how to improve FSQA?} \label{app:how_to_improve_FSQA}
\begin{figure}[t]
  \centering
  \includegraphics[width=0.6\linewidth]{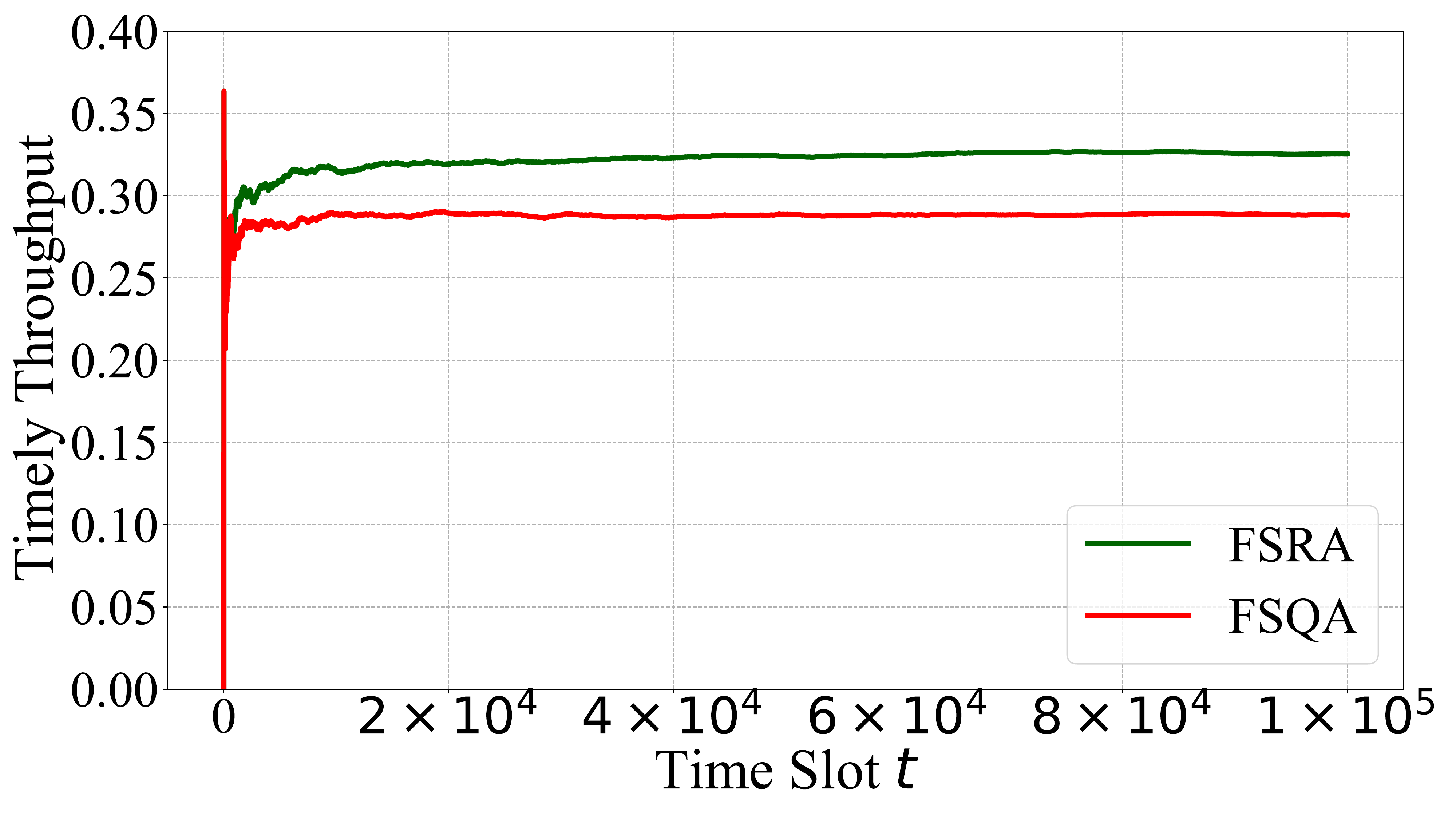}
  \caption{Compare the timely throughput of FSRA and FSQA where $p_b=0.5$, $p_b'=0.4$, $p_s=0.7$, $p_s'=0.6$, $p_t=0.4$, $D=2$.}\label{fig:performance_ql_rl}
\end{figure}

As we showed in Fig.~\ref{fig:compare_rl_ql}, the Q-learning-based FSQA algorithm
is worse than the R-learning-based FSRA. In this part, we consider a specific example for the two-device problem
to delve into the details of FSQA and FSRA.
We set system parameter settings as $p_b=0.5$, $p_b'=0.4$, $p_s=0.7$, $p_s'=0.6$, $p_t=0.4$, $D=2$.
The achieved system timely throughput of FSQA and FSRA is show in Fig.~\ref{fig:performance_ql_rl}.
Obviously, FSRA outperforms FSQA. Now we take a further step to examine the random access policies
of FSQA and FSRA, which are shown in Table~\ref{tab:FSRA-FSQA-policy}. As we can see, indeed, after convergence,
FSQA ahd FSRA take different policies, which thus results in different system timely throughput.

It is not clear which policy is better. We then use the upper-bound policy as a benchmark, i.e., \eqref{equ:upper-bound-policy},
to justify that the policy of FSRA is better. Note that in the upper-bound algorithm, we
use a model-based MDP formulation where Device 2 is aware of Device 1's queue information and parameters.
Thus, different from FSQA and FSRA whose system state is $s=(l_2,o)$ as shown in \eqref{equ:equ-state-FSQA-and-FSRA},
the system state of the upper-bound algorithm also includes Device 1's queue information, i.e., $s=(l_1,l_2,o)$,
as shown in \eqref{equ:S1}.  The policy of the upper-bound algorithm is shown in Table~\ref{tab:upper-bound-policy}.
Each state of FSQA and FSRA, i.e., $s=(l_2,o)$, corresponds to four states of upper-bound policy, i.e, $s=(l_1,l_2,o)$
where $l_1 \in \{(0,0),(0,1),(1,0),(1,1)\}$. For such four states of the upper-bound policy sharing the same $l_2$ and $o$,
we take a vote to obtain the majority action, which is the last column in Table~\ref{tab:upper-bound-policy}. The majority action
roughly represents the optimal action if the Device 2's queue information is $l_2$ and the channel observation is $o$.
We compare the majority action of the upper-bound policy in Table \ref{tab:upper-bound-policy} and the action of FSRA and FSQA in Table~\ref{tab:FSRA-FSQA-policy}.
We can see that FSRA has exactly the same action with the upper-bound policy for all states, while FSQA has different actions for
four states $(l_2,o)=((1,0), \textsf{SUCCESSFUL})$, $((1,1), \textsf{BUSY})$, $((1,1), \textsf{SUCCESSFUL}),$ and $((1,1), \textsf{FAILED})$.
With the help of the model-based upper-bound policy as a benchmark, we can see that indeed the policy of R-learning-based FSRA is better than
the policy of Q-learning-based FSQA.

\begin{figure}[t]
  \centering
  \includegraphics[width=0.6\linewidth]{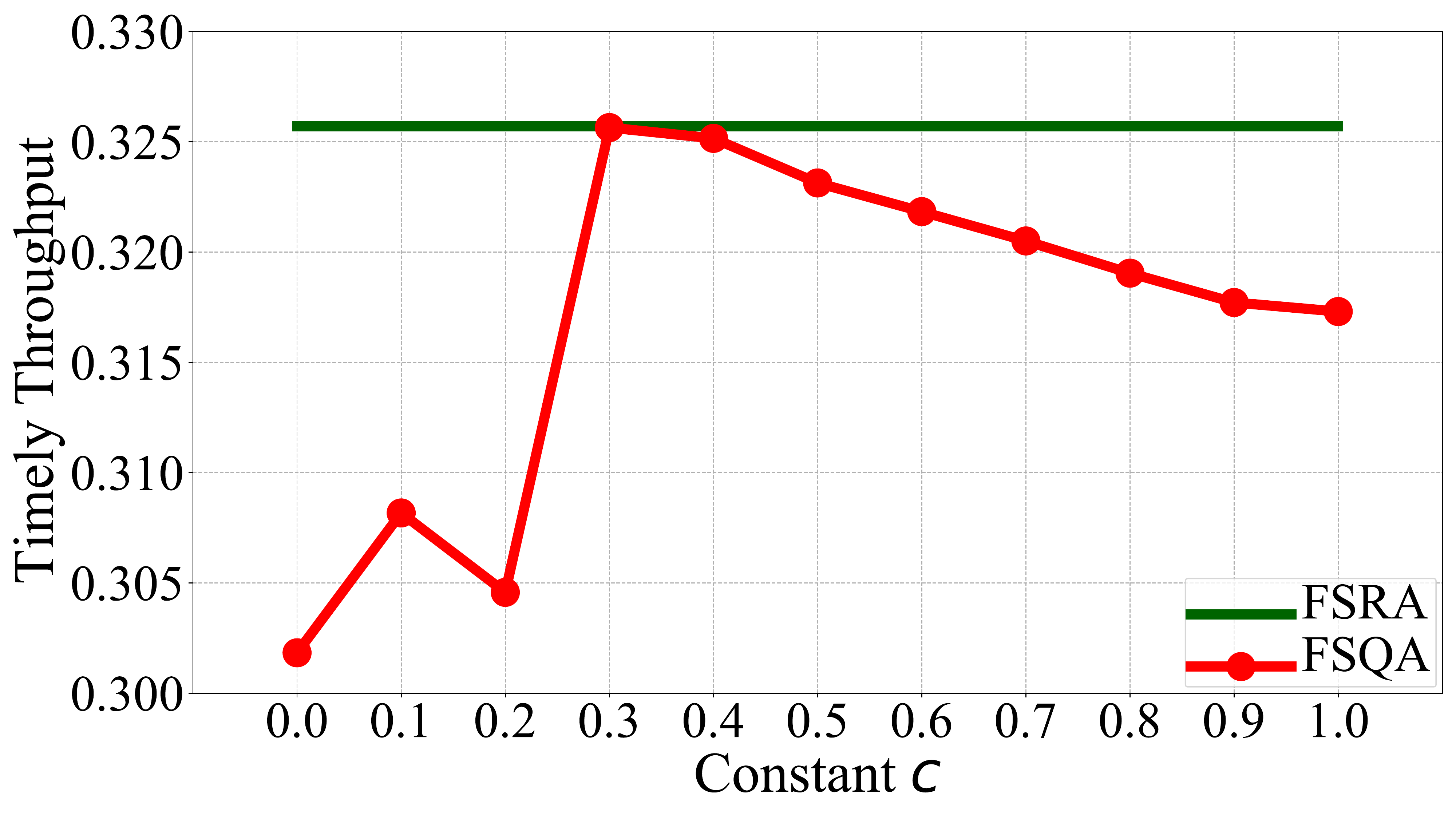}
  \caption{Compare the timely throughput of FSRA and the improved FSQA where $p_b=0.5$, $p_b'=0.4$, $p_s=0.7$, $p_s'=0.6$, $p_t=0.4$, $D=2$.}\label{fig:constant}
\end{figure}

Furthermore, we also use this example to show how to improve the performance of Q-learning-based FSQA algorithm.
Comparing the Q-function update of Q-learning in \eqref{equ:Q-learning-Q-function} and
the Q-function update of R-learning in \eqref{equ:upgrade_Q}, we can see that
the major difference is the parameter $\rho$.
Comparing \eqref{equ:Q-function-approx} and \eqref{equ:Q-approx-R-learning},
which respectively represents the physical meaning of Q-function for Q-learning and R-learning,
we can also observe that for average-reward MDP, we should use a relative value to response the reward. Namely,
the reward should be deducted by a constant $\rho$. To improve the Q-learning-based FSQA algorithm,
we thus re-define its reward function in \eqref{equ:reward function_RL} as
\bee
r(s_t,a_t) \triangleq 1_{\left\{o_t \in \{ \textsf{BUSY}, \textsf{SUCCESSFUL}\} \right \}}-c, \forall s_t \in \mathcal{S}', a_t \in \mathcal{A},
\label{equ:reward function_RL-redefined}
\eee
where $c \in [0,1]$ is a constant. We then compare the performance of FSRA and the improved FSQA algorithms with different $c$'s, as shown
in Fig.~\ref{fig:constant}. As we can see, when constant $c=0.3$, the improved FSQA achieves almost the same system timely throughput
with FSRA, which is much better the original FSQA algorithm (with $c=0$). In fact, parameter $\rho$ in \eqref{equ:upgrade_rho} of FSRA converges to 0.379 in this example.
Thus, the optimal constant $c=0.3$ in the improved FSQA is close to the converged $\rho$ of FSRA.
Although we can improve FSQA by re-defining its reward function according to \eqref{equ:reward function_RL-redefined},
there is generally no guidance on how to choose the best constant $c$, which is different for different problem instances.
Instead, in R-learning-based FSRA, the parameter $\rho$ is algorithmically  adjusted according to \eqref{equ:upgrade_rho} until its convergence.
This further demonstrates the benefit of R-learning over Q-learning for our studied problem.

\begin{table}[h]
\begin{center}
\caption{The policies of FSRA and FSQA after convergence when $p_b=0.5$, $p_b'=0.4$, $p_s=0.7$, $p_s'=0.6$, $p_t=0.4$ and $D=2$.
Note that channel observation $o=\text{B}$ means that $o=\textsf{BUSY}$,
$o=\text{S}$ means that $o=\textsf{SUCCESSFUL}$, $o=\text{I}$ means that $o=\textsf{IDLE}$,
and $o=\text{F}$ means that $o=\textsf{FAILED}$. \label{tab:FSRA-FSQA-policy}}
 \scriptsize
\begin{tabular}{|c|c|c|c|}
\hline
\multicolumn{2}{|c|}{State}     & FSRA         & FSQA         \\ \hline
\multicolumn{2}{|c|}{$s=(l_2, o)$} & \multicolumn{2}{c|}{action} \\ \hline
$l_2$               & $o$            & \multicolumn{2}{c|}{$a$}      \\ \hline
(0,0)            & B            & \textsf{WAIT}         & \textsf{WAIT}         \\ \hline
(0,0)            & S            & \textsf{WAIT}         & \textsf{WAIT}         \\ \hline
(0,0)            & I            & \textsf{WAIT}         & \textsf{WAIT}         \\ \hline
(0,0)            & F            & \textsf{WAIT}         & \textsf{WAIT}         \\ \hline
(0,1)            & B            & \textsf{TRANSMIT}     & \textsf{TRANSMIT}     \\ \hline
(0,1)            & S            & \textsf{TRANSMIT}     & \textsf{TRANSMIT}     \\ \hline
(0,1)            & I            & \textsf{TRANSMIT}     & \textsf{TRANSMIT}     \\ \hline
(0,1)            & F            & \textsf{TRANSMIT}     & \textsf{TRANSMIT}     \\ \hline
(1,0)            & B            & \textsf{TRANSMIT}     & \textsf{TRANSMIT}     \\ \hline
(1,0)            & S            & \textsf{TRANSMIT}     & \textsf{WAIT}         \\ \hline
(1,0)            & I            & \textsf{TRANSMIT}     & \textsf{TRANSMIT}     \\ \hline
(1,0)            & F            & \textsf{TRANSMIT}     & \textsf{TRANSMIT}     \\ \hline
(1,1)            & B            & \textsf{TRANSMIT}     & \textsf{WAIT}         \\ \hline
(1,1)            & S            & \textsf{TRANSMIT}     & \textsf{WAIT}         \\ \hline
(1,1)            & I            & \textsf{TRANSMIT}     & \textsf{TRANSMIT}     \\ \hline
(1,1)            & F            & \textsf{TRANSMIT}     & \textsf{WAIT}         \\ \hline
\end{tabular}
\end{center}
\end{table}

\begin{table}[]
\begin{center}
\caption{The upper-bound policy \eqref{equ:upper-bound-policy} when $p_b=0.5$, $p_b'=0.4$, $p_s=0.7$, $p_s'=0.6$, $p_t=0.4$ and $D=2$.
Note that channel observation $o=\text{B}$ means that $o=\textsf{BUSY}$,
$o=\text{S}$ means that $o=\textsf{SUCCESSFUL}$, $o=\text{I}$ means that $o=\textsf{IDLE}$,
and $o=\text{F}$ means that $o=\textsf{FAILED}$.\label{tab:upper-bound-policy}}
\renewcommand{\arraystretch}{0.65}
 \scriptsize
\begin{tabular}{|c|c|c|c|c|c|}
\hline
\multicolumn{3}{|c|}{State}                         & \multicolumn{2}{c|}{$\pi(s|a)$} & \multirow{3}{*}{Majority} \\ \cline{1-5}
\multicolumn{3}{|c|}{$s=(l_1, l_2, o)$}                 & \multicolumn{2}{c|}{$a$}       &                           \\ \cline{1-5}
$l_1$    & $l_2$                     & $o$                  & \textsf{WAIT}         & \textsf{TRANSMIT}      &                           \\ \hline
(0,0) & \multirow{4}{*}{(0,0)} & \multirow{4}{*}{B} & 1            & 0             & \multirow{4}{*}{\textsf{WAIT}}     \\ \cline{1-1} \cline{4-5}
(0,1) &                        &                    & 1            & 0             &                           \\ \cline{1-1} \cline{4-5}
(1,0) &                        &                    & 1            & 0             &                           \\ \cline{1-1} \cline{4-5}
(1,1) &                        &                    & 1            & 0             &                           \\ \hline
(0,0) & \multirow{4}{*}{(0,0)} & \multirow{4}{*}{S} & 1            & 0             & \multirow{4}{*}{\textsf{WAIT}}     \\ \cline{1-1} \cline{4-5}
(0,1) &                        &                    & 1            & 0             &                           \\ \cline{1-1} \cline{4-5}
(1,0) &                        &                    & 1            & 0             &                           \\ \cline{1-1} \cline{4-5}
(1,1) &                        &                    & 1            & 0             &                           \\ \hline
(0,0) & \multirow{4}{*}{(0,0)} & \multirow{4}{*}{I} & 1            & 0             & \multirow{4}{*}{\textsf{WAIT}}     \\ \cline{1-1} \cline{4-5}
(0,1) &                        &                    & 1            & 0             &                           \\ \cline{1-1} \cline{4-5}
(1,0) &                        &                    & 1            & 0             &                           \\ \cline{1-1} \cline{4-5}
(1,1) &                        &                    & 1            & 0             &                           \\ \hline
(0,0) & \multirow{4}{*}{(0,0)} & \multirow{4}{*}{F} & 1            & 0             & \multirow{4}{*}{\textsf{WAIT}}     \\ \cline{1-1} \cline{4-5}
(0,1) &                        &                    & 1            & 0             &                           \\ \cline{1-1} \cline{4-5}
(1,0) &                        &                    & 1            & 0             &                           \\ \cline{1-1} \cline{4-5}
(1,1) &                        &                    & 1            & 0             &                           \\ \hline
(0,0) & \multirow{4}{*}{(0,1)} & \multirow{4}{*}{B} & 0            & 1             & \multirow{4}{*}{\textsf{TRANSMIT}} \\ \cline{1-1} \cline{4-5}
(0,1) &                        &                    & 0            & 1             &                           \\ \cline{1-1} \cline{4-5}
(1,0) &                        &                    & 1            & 0             &                           \\ \cline{1-1} \cline{4-5}
(1,1) &                        &                    & 0            & 1             &                           \\ \hline
(0,0) & \multirow{4}{*}{(0,1)} & \multirow{4}{*}{S} & 0            & 1             & \multirow{4}{*}{\textsf{TRANSMIT}} \\ \cline{1-1} \cline{4-5}
(0,1) &                        &                    & 0            & 1             &                           \\ \cline{1-1} \cline{4-5}
(1,0) &                        &                    & 1            & 0             &                           \\ \cline{1-1} \cline{4-5}
(1,1) &                        &                    & 0            & 1             &                           \\ \hline
(0,0) & \multirow{4}{*}{(0,1)} & \multirow{4}{*}{I} & 0            & 1             & \multirow{4}{*}{\textsf{TRANSMIT}} \\ \cline{1-1} \cline{4-5}
(0,1) &                        &                    & 0            & 1             &                           \\ \cline{1-1} \cline{4-5}
(1,0) &                        &                    & 1            & 0             &                           \\ \cline{1-1} \cline{4-5}
(1,1) &                        &                    & 0            & 1             &                           \\ \hline
(0,0) & \multirow{4}{*}{(0,1)} & \multirow{4}{*}{F} & 0            & 1             & \multirow{4}{*}{\textsf{TRANSMIT}} \\ \cline{1-1} \cline{4-5}
(0,1) &                        &                    & 0            & 1             &                           \\ \cline{1-1} \cline{4-5}
(1,0) &                        &                    & 1            & 0             &                           \\ \cline{1-1} \cline{4-5}
(1,1) &                        &                    & 0            & 1             &                           \\ \hline
(0,0) & \multirow{4}{*}{(1,0)} & \multirow{4}{*}{B} & 0            & 1             & \multirow{4}{*}{\textsf{TRANSMIT}} \\ \cline{1-1} \cline{4-5}
(0,1) &                        &                    & 0            & 1             &                           \\ \cline{1-1} \cline{4-5}
(1,0) &                        &                    & 0.47916      & 0.52084       &                           \\ \cline{1-1} \cline{4-5}
(1,1) &                        &                    & 0.49551      & 0.50449       &                           \\ \hline
(0,0) & \multirow{4}{*}{(1,0)} & \multirow{4}{*}{S} & 0            & 1             & \multirow{4}{*}{\textsf{TRANSMIT}} \\ \cline{1-1} \cline{4-5}
(0,1) &                        &                    & 0            & 1             &                           \\ \cline{1-1} \cline{4-5}
(1,0) &                        &                    & 0            & 1             &                           \\ \cline{1-1} \cline{4-5}
(1,1) &                        &                    & 0            & 1             &                           \\ \hline
(0,0) & \multirow{4}{*}{(1,0)} & \multirow{4}{*}{I} & 0            & 1             & \multirow{4}{*}{\textsf{TRANSMIT}} \\ \cline{1-1} \cline{4-5}
(0,1) &                        &                    & 0            & 1             &                           \\ \cline{1-1} \cline{4-5}
(1,0) &                        &                    & 0.35845      & 0.64155       &                           \\ \cline{1-1} \cline{4-5}
(1,1) &                        &                    & 0.43222      & 0.56778       &                           \\ \hline
(0,0) & \multirow{4}{*}{(1,0)} & \multirow{4}{*}{F} & 0            & 1             & \multirow{4}{*}{\textsf{TRANSMIT}} \\ \cline{1-1} \cline{4-5}
(0,1) &                        &                    & 0            & 1             &                           \\ \cline{1-1} \cline{4-5}
(1,0) &                        &                    & 0            & 1             &                           \\ \cline{1-1} \cline{4-5}
(1,1) &                        &                    & 0            & 1             &                           \\ \hline
(0,0) & \multirow{4}{*}{(1,1)} & \multirow{4}{*}{B} & 0            & 1             & \multirow{4}{*}{\textsf{TRANSMIT}} \\ \cline{1-1} \cline{4-5}
(0,1) &                        &                    & 0            & 1             &                           \\ \cline{1-1} \cline{4-5}
(1,0) &                        &                    & 0.49072      & 0.50928       &                           \\ \cline{1-1} \cline{4-5}
(1,1) &                        &                    & 0.49713      & 0.50287       &                           \\ \hline
(0,0) & \multirow{4}{*}{(1,1)} & \multirow{4}{*}{S} & 0            & 1             & \multirow{4}{*}{\textsf{TRANSMIT}} \\ \cline{1-1} \cline{4-5}
(0,1) &                        &                    & 0            & 1             &                           \\ \cline{1-1} \cline{4-5}
(1,0) &                        &                    & 0            & 1             &                           \\ \cline{1-1} \cline{4-5}
(1,1) &                        &                    & 0            & 1             &                           \\ \hline
(0,0) & \multirow{4}{*}{(1,1)} & \multirow{4}{*}{I} & 0            & 1             & \multirow{4}{*}{\textsf{TRANSMIT}} \\ \cline{1-1} \cline{4-5}
(0,1) &                        &                    & 0            & 1             &                           \\ \cline{1-1} \cline{4-5}
(1,0) &                        &                    & 0.41202      & 0.58798       &                           \\ \cline{1-1} \cline{4-5}
(1,1) &                        &                    & 0.44799      & 0.55201       &                           \\ \hline
(0,0) & \multirow{4}{*}{(1,1)} & \multirow{4}{*}{F} & 0            & 1             & \multirow{4}{*}{\textsf{TRANSMIT}} \\ \cline{1-1} \cline{4-5}
(0,1) &                        &                    & 0            & 1             &                           \\ \cline{1-1} \cline{4-5}
(1,0) &                        &                    & 0            & 1             &                           \\ \cline{1-1} \cline{4-5}
(1,1) &                        &                    & 0            & 1             &                           \\ \hline
\end{tabular}
\end{center}
\end{table}

\end{appendix}


\end{document}